\documentclass[conference]{IEEEtran}

\newif\iflong
\longtrue 

\usepackage{cite}
\usepackage{amsmath,amssymb,amsfonts}
\usepackage{algorithmic}
\usepackage{graphicx}
\usepackage{textcomp}
\usepackage{placeins}
\usepackage{xcolor}
\usepackage{flushend}

\usepackage[sort,numbers,compress]{natbib}

\def\BibTeX{{\rm B\kern-.05em{\sc i\kern-.025em b}\kern-.08em
    T\kern-.1667em\lower.7ex\hbox{E}\kern-.125emX}}

\newcommand{\PROPOSE}{{\tt PROPOSE}}

\newcommand{\ACCEPT}{{\tt ACCEPT}}
\newcommand{\DELIVER}{{\tt DELIVER}}

\newcommand{\MULTICAST}{{\tt MULTICAST}}
\newcommand{\ACCEPTACK}{{\tt ACCEPT\_ACK}}
\newcommand{\NEWLEADER}{{\tt NEWLEADER}}
\newcommand{\NEWLEADERACK}{{\tt NEWLEADER\_ACK}}
\newcommand{\NEWSTATE}{{\tt NEW\_STATE}}
\newcommand{\NEWSTATEACK}{{\tt NEWSTATE\_ACK}}

\newcommand{\ttime}{{\sf time}}

\newcommand{\localts}{{\sf LocalTS}}
\newcommand{\globalts}{{\sf GlobalTS}}
\newcommand{\lastgts}{{\sf max\_delivered\_gts}}

\newcommand{\precond}{{\bf pre: }}

\newcommand{\clock}{{\sf clock}}

\newcommand{\lts}{\mathit{lts}}

\newcommand{\gts}{\mathit{gts}}

\newcommand{\ts}{t}
\newcommand{\leader}{{\sf leader}}
\newcommand{\ballot}{{\sf ballot}}
\newcommand{\aballot}{{\sf cballot}}
\newcommand{\vaballot}{\mathit{cballot}}

\newcommand{\Bal}{\mathit{Bal}}
\newcommand{\status}{{\sf status}}

\newcommand{\delivered}{{\sf Delivered}}
\newcommand{\phase}{{\sf Phase}}

\newcommand{\vphase}{\mathit{Phase}}
\newcommand{\vlocalts}{\mathit{LocalTS}}
\newcommand{\Lts}{\mathit{Lts}}
\newcommand{\vglobalts}{\mathit{GlobalTS}}

\newcommand{\vclock}{\mathit{clock}}
\newcommand{\dest}{{\sf dest}}

\newcommand{\Committed}{\textsc{committed}}
\newcommand{\Start}{\textsc{start}}
\newcommand{\Accepted}{\textsc{accepted}}
\newcommand{\Proposed}{\textsc{proposed}}

\newcommand{\LEADER}{\textsc{leader}}
\newcommand{\FOLLOWER}{\textsc{follower}}

\newcommand{\RECOVERING}{\textsc{recovering}}

\newcommand{\false}{\textsc{false}}
\newcommand{\true}{\textsc{true}}
\newcommand{\GST}{\mathsf{GST}}
\newcommand{\G}{\mathcal{G}}
\newcommand{\A}{\mathcal{A}}
\newcommand{\Q}{\mathcal{Q}}
\newcommand{\Leads}{\mathcal{L}}
\newcommand{\Proc}{\mathcal{P}}
\newcommand{\M}{\mathcal{M}}
\newcommand{\multicast}{\texttt{multicast}}
\newcommand{\deliverapp}{\texttt{deliver}}
\newcommand{\retry}{\texttt{retry}}
\newcommand{\currleaders}{\mathsf{Cur\_leader}}

\newcounter{saveenum}

\makeatletter
\newcommand{\removelatexerror}{\let\@latex@error\@gobble}
\makeatother

\usepackage[vlined,linesnumbered,noresetcount]{algorithm2e}
\SetKwBlock{SubAlgoBlock}{}{end}
\newcommand{\SubAlgo}[2]{#1 \SubAlgoBlock{#2}}
\let\oldnl\nl
\newcommand{\nonl}{\renewcommand{\nl}{\let\nl\oldnl}}

\makeatletter
\renewcommand\@IEEEsectpunct{}
\renewcommand\paragraph{\@startsection{paragraph}{5}{\z@}%
                                       {2ex \@plus1ex \@minus .2ex}%
                                       {-1em}%
                                      {\normalsize\bfseries}}
\makeatother

\newcommand{\tr}[2]{\iflong{}\S#1\else{}\cite[\S#2]{ext}\fi}

\newcommand{\ncorrectness}{A}
\newcommand{\nexp}{B}

\newtheorem{theorem}{Theorem}

\newtheorem{lemma}{Lemma}

\def\squareforqed{\hbox{\rlap{$\sqcap$}$\sqcup$}}
\def\qed{\ifmmode\squareforqed\else{\unskip\nobreak\hfil
\penalty50\hskip1em\null\nobreak\hfil\squareforqed
\parfillskip=0pt\finalhyphendemerits=0\endgraf}\fi}

\begin{document}

\SetKw{WhenReceived}{when received}
\SetKw{WereReceived}{were received}
\SetKw{Var}{var}
\SetKw{Send}{send}
\SetKw{KwFrom}{from}
\SetKw{KwTo}{to}
\SetKw{FromQuorum}{from a quorum of}
\SetKw{OnceReplies}{once replies}
\SetKw{FromClient}{from a client or}
\SetKw{Including}{including}
\SetKw{Excluding}{excluding}
\SetKw{KwWhere}{where}
\SetKw{Fun}{function}

\title{White-Box Atomic Multicast}

\author{
\IEEEauthorblockN{Alexey Gotsman} 
\IEEEauthorblockA{IMDEA Software Institute}
\and
\IEEEauthorblockN{Anatole Lefort}
 \IEEEauthorblockA{T\'el\'ecom SudParis}
\and 
 \IEEEauthorblockN{Gregory Chockler} 
 \IEEEauthorblockA{Royal Holloway, University of London}
}

\maketitle

\begin{abstract}
  Atomic multicast is a communication primitive that delivers messages to
  multiple groups of processes according to some total order, with each group
  receiving the projection of the total order onto messages addressed to it. To
  be scalable, atomic multicast needs to be genuine, meaning that only the
  destination processes of a message should participate in ordering it. In this
  paper we propose a novel genuine atomic multicast protocol that in the absence
  of failures takes as low as 3 message delays to deliver a message when no
  other messages are multicast concurrently to its destination groups, and 5
  message delays in the presence of concurrency. This improves the latencies of
  both the fault-tolerant version of classical Skeen's multicast protocol (6 or
  12 message delays, depending on concurrency) and its recent improvement by
  Coelho et al. (4 or 8 message delays).  To achieve such low latencies, we
  depart from the typical way of guaranteeing fault-tolerance by replicating
  each group with Paxos. Instead, we weave Paxos and Skeen's protocol together
  into a single coherent protocol, exploiting opportunities for white-box
  optimisations. We experimentally demonstrate that the superior theoretical
  characteristics of our protocol are reflected in practical performance
  pay-offs.
\end{abstract}

\section{Introduction}
\label{sec:intro}

Machine crashes are a fact of life in modern cloud services. The classical way
of enabling the services to tolerate such failures is using a state-machine
replication approach~\cite{smr}: a service is defined by a deterministic state
machine and is run on several replicas, each maintaining its own local copy of
the machine. Different copies can be kept in sync using an {\em atomic
  broadcast} protocol, which delivers {\em application messages} to replicas in
some total order and thereby ensures that they evolve in the same
way. Unfortunately, it is often impossible for a single machine to store the
whole service state. A solution is to partition the service across several {\em
  process groups}, each containing several replicas to guarantee
fault-tolerance. In this setting, replica consistency can be maintained using
{\em atomic multicast}~\cite{broadcast-survey}. This accepts application
messages together with sets of groups they are relevant to and delivers messages
to their destination groups according to some total order, so that each group
receives the projection of the total order onto messages addressed to it
(\S\ref{sec:problem}). Atomic multicast thus generalises atomic broadcast, since
it provides the same guarantees in the case when there is a single process
group. 

Ideally, we want an atomic multicast protocol to be {\em genuine}, i.e.,
only the processes in the destination groups of a message should participate in
ordering it~\cite{Guerraoui2001}. This allows messages to disjoint sets of
groups to be ordered in parallel, thus enabling scalability. For example,
genuine atomic multicast has been used to scale fault-tolerant transaction
processing systems~\cite{granola,pstore,nmsi} and log-based
systems~\cite{mahesh}. Genuine atomic multicast essentially requires
constructing a total order on application messages addressed to different groups
in a decentralised way. Achieving this is challenging, and 
classical implementations of genuine atomic multicast have suboptimal
performance. In this paper we set out to improve this situation.

The most well-known protocol for atomic multicast is folklore Skeen's protocol
(described, e.g., in~\cite{Guerraoui2001}), which handles a restricted setting
where each group consists of a single reliable process (\S\ref{sec:skeen}). In a
nutshell, the protocol creates a total order on application messages by
assigning them unique timestamps, computed similarly to Lamport
clocks~\cite{lamport-clock}. To multicast an application message, a client
process sends it to all its destinations. Each destination process generates a
{\em local timestamp} from a local logical clock and sends it to the other
destination processes. When a process receives all local timestamps for a given
message, it computes its final {\em global timestamp} as their maximum and
advances its clock to be no lower than the timestamp. A process can deliver an
application message once it is sure that no message will get a lower global
timestamp. If the process receives new messages while
waiting for this condition to be met, the process may need to delay the delivery of the
message past the point when its global timestamp is known; 
this phenomenon is known as a {\em convoy effect}~\cite{pierre-convoy}.
  For this reason, Skeen's protocol has the
latency of $2$ message delays when processing a solo message and $4$ message
delays when multiple messages are multicast concurrently to the same group. To
capture this difference in complexity, we introduce metrics, 
called {\em collision-free latency} and {\em failure-free latency}, that 
respectively bound the latency in the above two situations 
(and in the absence of failures).

A common approach to making Skeen's protocol
fault-tolerant~\cite{raynal,multicast-cost} is to get every group to simulate a
reliable process in Skeen's using a replication protocol, such as
Paxos~\cite{paxos}. In this case each of the two key actions of Skeen's
protocol---computing a local timestamp and advancing the clock above a global
timestamp---requires a round trip from the Paxos leader of each destination
group to a quorum of processes in the same group, to persist the effect of the
action. The resulting protocol has the collision-free and failure-free latencies
of $6$ and $12$ message delays, respectively; this is prohibitively high,
especially when multicast is used in a wide-area network.

In this paper we present a novel fault-tolerant atomic multicast protocol that
lowers the collision-free and failure-free latencies to $3$ and $5$ message
delays, respectively (\S\ref{sec:optimised}).
This improves on a recent optimised version of Skeen's protocol by Coelho et
al.~\cite{fastcast}, which has the collision-free and failure-free latencies of
$4$ and $8$ message delays. In particular, our protocol narrows the 2x gap
between the two metrics typical of existing atomic multicast implementations.

To achieve such low latencies, we depart from the standard designs of
fault-tolerant multicast protocols, which have used consensus as a black
box~\cite{raynal,multicast-cost,fastcast,scalcast}. Instead,  we combine the
ideas from Skeen's protocol with those of Paxos into a single coherent
protocol. This allows us to exploit several white-box optimisations that lead
to a more efficient solution.



In more detail, our protocol takes the {\em passive replication}
approach~\cite{zab,vr}: a special {\em leader} process in each group computes
the timestamps and decides when to deliver an application message like in
Skeen's protocol; the rest of the processes merely {\em follow} its
decisions. To replicate leader actions when multicasting an application message,
the protocol performs a message exchange similar to the one of Paxos, but
between all leaders of the destination groups on the one hand and majorities of
followers in all destination groups on the other. This message exchange
replicates both of the key actions of Skeen's protocol---assigning a local
timestamp and advancing the clock above the global timestamp---in a single round
trip, thus minimising delivery latency. Since in our protocol the leader takes
decisions about delivery unilaterally, based on its local state, every decision
it takes on a message only makes sense in the context of its previous decisions
on other messages. This requires care when recovering from a leader failure:
recovery cannot be done for each application message independently (like in
multi-Paxos~\cite{paxos}), but has to be done for all messages at once (like in
Viewstamped Replication~\cite{vr} and Zab~\cite{zab}). We rigorously prove that
our white-box protocol is correct (\S\ref{sec:correctness}\iflong{}
and~\tr{\ref{app:correctness}}{\ncorrectness}\fi). We also propose a method for
analysing the latency of Skeen-based protocols, which connects collision-free
and failure-free latencies and is applicable to both our protocol and previous
proposals.


Finally, we experimentally demonstrate that the superior theoretical
characteristics of our protocol are reflected in practical performance pay-offs
(\S\ref{sec:exp}). Our protocol outperforms the state-of-the-art protocol by
Coelho et al.~\cite{fastcast} on latency and throughput by 2x on average.



\section{Problem Statement}
\label{sec:problem}

We consider an asynchronous message-passing system consisting of a finite set
of $N$ processes $\Proc$, which can fail by crashing. A process is  {\em
correct}\/ if it never crashes, and {\em faulty}\/
otherwise. Processes are connected by reliable FIFO channels, i.e., messages
are delivered in the FIFO order, and every message sent by a process $p$ to
another process $q$ is guaranteed to be eventually delivered by $q$ provided
both $p$ and $q$ are correct.

We fix $\G \in 2^\Proc$ to be a set of {\em process groups} and let $|\G|=k$. We
assume that the process groups are disjoint, i.e.,
$\forall g_1, g_2 \in \G.\, g_1 \cap g_2 = \emptyset$. Every group $g \in \G$
consists of $2f +1$ processes, at most $f$ of which can fail. We call a set
of $f+1$ processes in $g$ a {\em quorum} in $g$. The assumption of disjoint
groups is standard for practical multicast
protocols~\cite{raynal,scalcast,fastcast}. It captures common usage scenarios in
which atomic multicast is deployed for replicating a partitioned data
store~\cite{granola,pstore,nmsi}, and it does not prevent collocating processes
that are members of different groups on the same machine.



We consider the problem of implementing atomic multicast in the above system,
which allows a process to send an {\em application message} $m$ from a set
$\M$ to a set of {\em destination groups} $\dest(m) \subseteq \G$. We denote
the events of multicasting a message $m$ and delivering it by $\multicast(m)$
and $\deliverapp(m)$, respectively. For simplicity, we assume that all
messages multicast in a single execution are unique.   
A message $m$ is {\em partially delivered} if it is delivered by some
process in all its destination groups.
A message $m$ is {\em concurrent}\/ with a message $m'$ if 
$m'$ is multicast before $m$ is partially delivered,
and $m$ is multicast before $m'$ is partially delivered.
Two
messages $m$ and $m'$ are {\em conflicting}\/ if $\dest(m) \cap \dest(m') \neq
\emptyset$.


An algorithm is a correct implementation of atomic multicast if its every run
satisfies the following:
\begin{itemize}
\item \textbf{Validity.} If a process in a group $g$ delivers a message $m$, then some process has multicast $m$ before and $g \in \dest(m)$.

\item \textbf{Integrity.} Every process delivers a message at most once.
  
\item \textbf{Ordering.} There exists a total order $\prec$ on the set of all
  messages multicast in the run such that, if a process $p$ delivers $m$, then
  for all messages $m' \prec m$, $p$ delivers $m'$ before $m$ provided
  $p \in g$ for some $g\in \dest(m')$.

\item {\bf Termination.} For every message $m$, if $m$ is either multicast by a
  correct process or delivered by any process, then for all groups
  $g\in \dest(m)$, $m$ is eventually delivered by a quorum of processes in $g$.

%
\end{itemize}
In particular, the ordering property ensures that each group receives the
projection of a single total order onto messages addressed to it.



  

A protocol implementing atomic multicast is {\em
  genuine}~\cite{Guerraoui2001,raynal} if it satisfies the following {\em
  minimality}\/ property in every run: if $m$ is multicast in the run, then for
every process $p$ that participates in ordering $m$, the process $p$ is either
$m$'s sender or a member of some $g\in \dest(m)$.

By instantiating atomic multicast with a single group comprising all processes
in $\Proc$ we get {\em atomic broadcast}~\cite{HT94}, which delivers messages to
all processes. Since atomic broadcast is equivalent to consensus~\cite{CT96}, it
cannot be implemented in an asynchronous environment with
failures~\cite{FLP85}. To circumvent this impossibility, we assume that the
system eventually becomes {\em failure-free}, i.e., the process failures cease
to occur and message delays are upper-bounded by an a priori fixed constant
$\delta$. {\em Global stabilization time} ($\GST$)~\cite{DLS88} is the time
(unknown to the algorithm) such that the onset of a failure-free period is
guaranteed to occur no later than at $\GST$ in every run.


To measure time complexity of an atomic multicast implementation, we assign
every event in a run a non-decreasing real-valued time such that after $\GST$,
the time elapsing between every pair of matching send and receive events of a
protocol message is at most $\delta$, and every step executed locally by a
process is instantaneous. For a message $m$ multicast in a run, 
and a group $g\in \dest(m)$, $m$'s {\em delivery latency}\/ with 
respect to $g$
is the time elapsing between $\multicast(m)$ and the
earliest $\deliverapp(m)$ by some process in $g$. 
An atomic multicast protocol has a
{\em failure-free latency} of $\Delta$ if for every run there exists a time
$t \ge \GST$ such that for every application message $m$ multicast after $t$,
$m$'s delivery latency is at most $\Delta$ with respect to all groups
in $\dest(m)$. A protocol has a 
{\em collision-free latency} of $\Delta$ if for every run, 
there exists a time $t \ge \GST$ such
that for every application message $m$ multicast after $t$ that does not
conflict with any concurrent messages multicast by correct processes, 
$m$'s delivery latency is at most
$\Delta$ with respect to all groups in $\dest(g)$. 
Note that our latency metrics are computed based on the first delivery
of a message in every destination group, 
whereas metrics used in previous work use the last
one~\cite{multicast-cost}. Our choice more faithfully reflects the
client-perceived latency in practical use cases of multicast, where the first
process that delivers a message can process it and reply to the
client~\cite{granola,pstore,nmsi}.










\section{Skeen's Protocol}
\label{sec:skeen}

\begin{figure}
\scalebox{0.98}{%
\removelatexerror
\begin{algorithm}[H]
  \setcounter{AlgoLine}{0}

$\clock \leftarrow 0 \in \mathbb{N}$;
\\
$\phase[\,] \leftarrow (\lambda k.\, \Start) \in {}$
\\
\nonl $\phantom{\phase[\,] \leftarrow {}} 
(\M \to \{\Start, \Proposed, \Committed\})$;\!
\\
$\localts[\,] \in \M \to (\mathbb{N} \times \G)$;
\\
$\globalts[\,] \in \M \to (\mathbb{N} \times \G)$;
\\
$\delivered \leftarrow (\lambda k.\, \false) \in \M \to \{\false,
\true\}$

  \smallskip
  \smallskip

  \SubAlgo{${\tt multicast}(m)$\label{sk:client}}{
    \Send $\MULTICAST(m)$ {\bf to} $\dest(m)$\;
  }

  \smallskip
  \smallskip

  \SubAlgo{\WhenReceived $\MULTICAST(m)$\label{sk:receive-msg}}{
    $\clock \leftarrow \clock+1$\; \label{sk:clock-inc}
    $\localts[m] \leftarrow (\clock, g_0)$\; \label{sk:origin}
    $\phase[m] \leftarrow \Proposed$\;\label{sk:set-phase-proposed}
    \Send $\PROPOSE(m, g_0, \localts[m])$ {\bf to} $\dest(m)$\;
  }

  \smallskip
  \smallskip

  \SubAlgo{\WhenReceived $\PROPOSE(m, g, \Lts(g))$ \\
    \nonl\quad {\bf for every} $g \in \dest(m)$\label{sk:receive-propose}}{
    $\globalts[m] \leftarrow \max\{\Lts(g) \mid g \in \dest(m)\}$\;\label{sk:store-gts}
    $\clock \leftarrow \max\{\clock, \ttime(\globalts[m])\}$\;\label{sk:clock-max} 
    $\phase[m] \leftarrow \Committed$\;
    \ForAll{$\begin{array}[t]{@{}l@{}l@{}}
               \{m' \mid {} & \phase[m'] = \Committed \wedge {} \\
               & \delivered[m'] = \false \wedge {} \\
               & \forall m''.\, \phase[m''] = \Proposed {\implies} \\
               & \phantom{\forall m''.\,} \localts[m''] > \globalts[m']\}
             \end{array}$\label{sk:deliv-check1}\\
      \nonl $\quad$\textbf{\em ordered by} $\globalts[m']$}{
      $\delivered[m'] \leftarrow \true$\;\label{sk:delivered-true}
      {\tt deliver}$(m')$;\label{sk:deliv-follower}
    }}
\end{algorithm}
}
\caption{Skeen's protocol at a process $p_i \in g_0$.}
\label{fig:skeen}
\end{figure}

We first consider an idealised setting where each group in $\G$ consists of a
single reliable process. In this setting, genuine atomic multicast can be
implemented using folklore Skeen's protocol (described, e.g.,
in~\cite{Guerraoui2001}). This protocol serves as a basis for our optimised
fault-tolerant protocol and, hence, we review it first. We give its pseudocode
in Figure~\ref{fig:skeen}.

The protocol creates a total order on application messages by assigning them
unique timestamps, computed similarly to Lamport
clocks~\cite{lamport-clock}. Timestamps are pairs $(t, g)$ of a non-negative
integer $t \in \mathbb{N}$ and a group identifier $g \in \G$. They are ordered
lexicographically using an arbitrary total order on $\G$, with a special
timestamp $\bot$ being the minimal timestamp. For a timestamp
$\mathit{ts} = (\ts, g)$ we let $\ttime(\mathit{ts}) = \ts$.

To multicast an application message $m$, a process sends it in a $\MULTICAST$
message to the destination groups $\dest(m)$ (line~\ref{sk:client}). Each
process maintains an integer $\clock$, used to generate timestamps. When a
process in a group $g_0$ receives $\MULTICAST(m)$ (line~\ref{sk:receive-msg}),
it increments the clock and computes a {\em local timestamp of $m$ at group
  $g_0$} as the pair of the resulting clock value and the group identifier
$g_0$. This timestamp can be viewed as $g_0$'s proposal of what the final
timestamp of $m$ should be; it is stored in a $\localts$ array\footnote{To aid
  understanding, in this paper we capitalise the names of arrays and
  vectors.}. The process keeps track of the status of application messages being
multicast in an array $\phase$, whose entries initially store $\Start$. When the
process computes a local timestamp for $m$, it advances $m$'s phase to
$\Proposed$. It then sends the local timestamp in a $\PROPOSE$ message to all
the destinations of $m$ (including itself, for uniformity).

A process that is a destination of $m$ acts once it receives a $\PROPOSE$
message for $m$ from each destination group $g \in \dest(m)$, which carries
$m$'s local timestamp $\Lts(g)$ at $g$ (line~\ref{sk:receive-propose}). The
process computes the final {\em global timestamp} of $m$ as the maximal of its
local timestamps and stores it in a $\globalts$ array. The process also advances
the phase of $m$ to $\Committed$ and ensures that its clock is no lower than the
first part of the global timestamp. Note that all destinations of $m$ will
receive the same sets of local timestamps for $m$ and will thus compute the same
global timestamp. Additionally, global timestamps are unique for each
application message: if two messages got the same global timestamp $(n, g)$,
then they must have got the same local timestamp from group $g$; but this is
impossible because a process increments its clock when issuing a local timestamp
(line~\ref{sk:clock-inc}).

Having computed the global timestamp for $m$, the process tries to deliver one
or more committed messages (line~\ref{sk:deliv-check1}). A Boolean array
$\delivered$ keeps track of whether a given message has been delivered. Messages
are delivered in the order of their global timestamps; hence, the process can
deliver a message $m'$ only if it has already delivered all messages addressed
to it with a lower global timestamp. A subtlety is that the process does not
know the global timestamps for the messages $m''$ that are in the $\Proposed$
phase. Hence, the process only delivers $m'$ if all such messages $m''$ have
local timestamps higher than the global timestamp of $m'$: then their global
timestamps will also be higher than that of $m'$. Note that this check is
complete: application messages the process will receive for multicasting after
delivering $m'$ will get global timestamps higher than $\globalts[m']$. This is
because, when the process commits $m'$, it advances its clock so that it is no
lower than $\globalts[m']$ (line~\ref{sk:clock-max}). Thus, any application
message the process receives afterwards will get a local timestamp at $g_0$
higher than $\globalts[m']$ and, thus, will also get a global timestamp higher
than $\globalts[m']$.

\smallskip
\smallskip

\begin{theorem}
  Skeen's protocol in Figure~\ref{fig:skeen} is a genuine implementation of
  atomic multicast among singleton groups.
\end{theorem}

\smallskip
\smallskip

Note that in Skeen's protocol a process can increase its clock at any time
without violating correctness. In \S\ref{sec:optimised} we use this insight to
construct a fast fault-tolerant version of this protocol.

\begin{figure}[t]
\begin{centering}
\includegraphics[width=0.9\columnwidth]{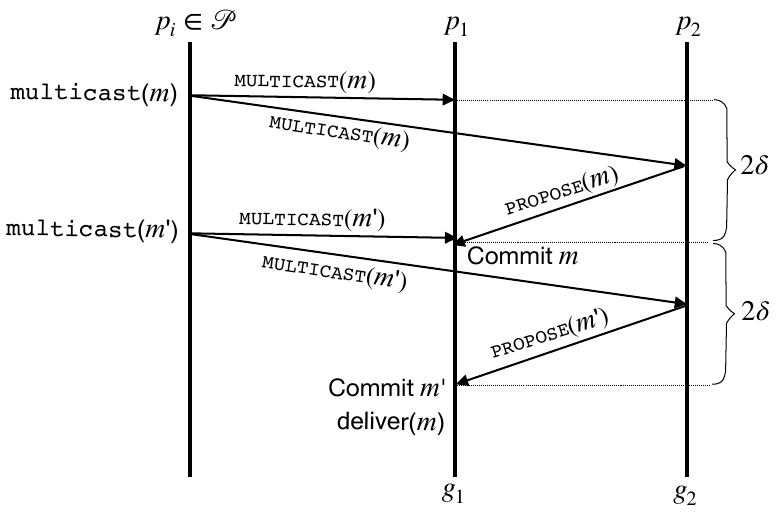}
\end{centering}
\caption{Message-flow diagram illustrating the convoy effect in Skeen's
protocol.}
\label{fig:convoy}
\end{figure}

Skeen's protocol has the collision-free latency of $2\delta$ ($\MULTICAST$,
$\PROPOSE$). However, its failure-free latency is higher because in this
protocol a committed message $m$ is blocked from delivery as long as there are
any uncommitted messages with a local timestamp lower than $m$'s global
timestamp. As a result, $m$'s delivery latency at a process $p_i$ may exceed
the collision-free latency of $2\delta$ in case an application message is
received before the $p_i$'s clock has been advanced past $m$'s global
timestamp---a phenomenon known as a {\em convoy effect}~\cite{pierre-convoy}.

The exact amount of extra delay depends on the timing of the arrival of a
conflicting message $m'$, and can, in the worst case, be as high as $2\delta$.
This is demonstrated by the scenario in Figure~\ref{fig:convoy}, where the
$\MULTICAST(m')$ message, triggered by $\multicast(m')$ with
$\dest(m') = \{g_1, g_2\}$, is received by $p_1$ immediately before $m$ is
committed at this process. Since $p_1$'s clock is still lower than
$\globalts[m]$ at the time $m'$ is received, this message is assigned a local
timestamp less than $\globalts[m]$. As a result, the delivery of $m$ must now be
delayed until $m'$ commits. In the worst-case scenario of
Figure~\ref{fig:convoy} this takes another $2\delta$, because $\MULTICAST(m')$
takes close to $0$ to arrive at $p_1$, but exactly $\delta$ to arrive at $p_2$;
then $\PROPOSE(m')$ from $p_2$ also takes exactly $\delta$ to arrive at
$p_1$. Thus, the failure-free latency of Skeen's protocol is in fact $4\delta$,
i.e., double its collision-free latency.



\section{White-Box Protocol}
\label{sec:optimised}

We now consider the general setting where each group consists of $2f+1$
processes, out of which at most $f$ can fail. A straightforward way to implement
atomic multicast in this setting is to use state-machine replication to make a
group simulate a reliable process in Skeen's protocol~\cite{raynal}; this is
usually based on a consensus protocol such as Paxos~\cite{paxos}. Then in
addition to $\MULTICAST$ and $\PROPOSE$ messages, the resulting protocol
requires two round trips from the Paxos leader of a group to a quorum of
processes in the same group---one to persist the local timestamp
(line~\ref{sk:origin} in Figure~\ref{fig:skeen}) and another to persist the
global timestamp and update the clock
(lines~\ref{sk:store-gts}-\ref{sk:clock-max}). Hence, as we show in
\S\ref{sec:correctness}, the resulting protocol has the collision-free latency
of $6\delta$ and failure-free latency is $12\delta$ (due to the convoy
effect). In this section we present a protocol that lowers the collision-free
latency to $3\delta$ and the failure-free latency to $5\delta$ by weaving
together Skeen's protocol across groups and a Paxos-like protocol within each
group.

\begin{figure}[t]
\includegraphics[width=\columnwidth]{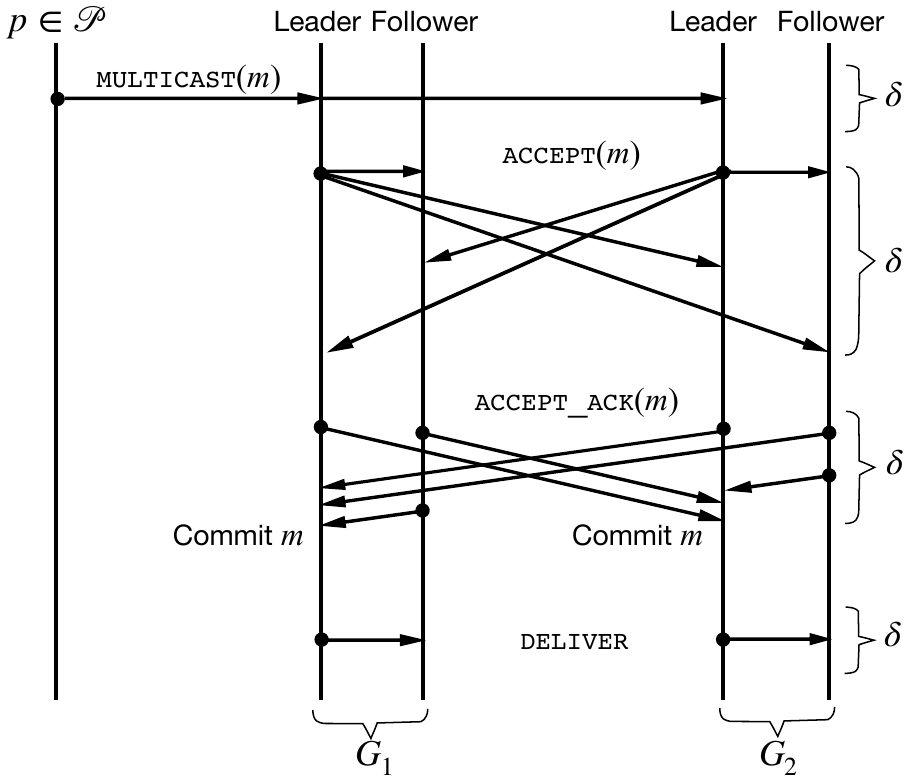}
\vspace*{-0.8cm}
\caption{Message-flow diagram illustrating the behaviour of the white-box
  protocol in a collision-free scenario. On the right-hand side
  we give the maximum time each protocol step can take.}
\label{fig:flow}
\end{figure}

We list the variables maintained by our protocol in Figure~\ref{fig:vars}, give
its pseudocode in Figure~\ref{fig:protocol}, illustrate the message flow of the
protocol in Figure~\ref{fig:flow} and summarise the key invariants used in its
proof of correctness in Figure~\ref{fig:inv}.

\begin{figure}[t]
\scalebox{0.98}{%
$\begin{array}{@{}l@{}}
  \clock \leftarrow 0 \in \mathbb{N}
  \\[1pt]
  \phase[\,] \leftarrow (\lambda k.\, \Start) \in {}
  \\[1pt]
  ~~~~~~~~~\,\M \to \{\Start, \Proposed, \Accepted, \Committed\}
  \\[1pt]
  \localts[\,] \in \M \to (\mathbb{N} \times \G)
  \\[1pt]
  \globalts[\,] \in \M \to (\mathbb{N} \times \G)
  \\[1pt]
  \delivered \leftarrow (\lambda k.\, \false) \in \M \to \{\false, \true\}
  \\[1pt]
  \status \in \{\LEADER,\FOLLOWER, \RECOVERING\}
  \\[1pt]
  \aballot \leftarrow \bot \in (\mathbb{N} \times \Proc) \cup \{\bot\}
  \\[1pt]
  \ballot \leftarrow \bot \in (\mathbb{N} \times \Proc) \cup \{\bot\}
  \\[1pt]
   \currleaders[\,] \in \G \to \Proc
  \\[1pt]
  \lastgts \leftarrow \bot \in (\mathbb{N} \times \G) \cup \{\bot\}
\end{array}$
}
\caption{Variables of a process in the white-box multicast protocol.}
\label{fig:vars}
\end{figure}


\begin{figure*}[p]
\begin{tabular}{@{}l@{\ \ }|@{\ \ \ \ }l@{}}
\begin{minipage}{9cm}
\scalebox{0.98}{%
\removelatexerror
\begin{algorithm}[H]
\setcounter{AlgoLine}{0}

  \SubAlgo{${\tt multicast}(m)$\label{alg:client}}{
    \mbox{\Send $\MULTICAST(m)$ {\bf to} $\{\currleaders[g] \mid g \in \dest(m)\}$\;}\label{alg:send-multicast}
  }

  \smallskip
  \smallskip
  \smallskip

  \SubAlgo{\WhenReceived $\MULTICAST(m)$\label{alg:leader-origin}}{
    \precond $\status = \LEADER$\;\label{alg:status-leader}
    \If{$\phase[m] = \Start$\label{alg:lts-exists}}{%
      $\clock \leftarrow \clock+1$\; \label{alg:clock-inc}
      $\localts[m] \leftarrow (\clock, g_0)$\; \label{alg:origin}
      $\phase[m] \leftarrow \Proposed$\;\label{alg:set-phase-proposed}
    }
    \mbox{\Send $\ACCEPT(m, g_0, \aballot, \localts[m])$ {\bf to} $\dest(m)$;}\label{alg:send-accept}
  }

  \smallskip
  \smallskip
  \smallskip

  \SubAlgo{\WhenReceived $\ACCEPT(m, g, \Bal(g), \Lts(g))$\label{alg:receive-accept}\\
    \nonl \quad {\bf for every} $g \in \dest(m)$}{ 
    \precond $\status \in \{\FOLLOWER, \LEADER\} \wedge {}$\label{alg:check-propose}
    \nonl $\phantom{\text{\precond}} \aballot = \Bal(g_0)$\;
    \If{$\phase[m] \in \{\Start, \Proposed\}$\label{alg:check-accept-commit}}{%
      $\phase[m] \leftarrow \Accepted$;\label{alg:set-accept-commit}}
    $\localts[m] \leftarrow \Lts(g_0)$\;\label{alg:store-lts}
    $\begin{array}[t]{@{}l@{}l@{}}
       \clock \leftarrow \max\{&\ttime(\max\{\Lts(g) \mid g \in \dest(m)\}),\\
                                 & \clock\};\end{array}$\label{alg:clock-max}\\
    \ForAll{$g \in \dest(m)$}{%
      \mbox{\Send $\ACCEPTACK(m, g_0, \Bal)$ \KwTo $\leader(\Bal(g))$;}}\label{alg:send-acceptack}
  }

  \smallskip
  \smallskip
  \smallskip

  \SubAlgo{\WhenReceived $\ACCEPTACK(m, g, \Bal)$\label{alg:receive-acceptack}\\
    \nonl \quad \FromQuorum $p_j \in g$ {\bf in each $g \in \dest(m)$}
    \\
    \nonl \quad {\bf including myself and previously received}
    \\
     \nonl \mbox{\quad $\ACCEPT(m, g, \Bal(g), \Lts(g))$ 
      {\bf for every} $g \in \dest(m)$}\label{alg:acceptack}}{
    \precond $\status = \LEADER \wedge \aballot = \Bal(g_0)$\; \label{alg:pre-acceptack}
    $\globalts[m] \leftarrow \max\{\Lts(g) \mid g \in \dest(m)\}$\;\label{alg:gts-max}
    $\phase[m] \leftarrow \Committed$\;\label{alg:set-committed-phase}
    \ForAll{%
      $\!\{m' \mid \phase[m'] = \Committed \wedge {}$\label{alg:deliv-check1}
      \\
      \nonl $\phantom{\mbox{{\bf forall} $\!\{ m' \mid {}$}} \delivered[m'] = \false \wedge {}$
      \\
      \nonl \mbox{$\phantom{\mbox{{\bf forall} $\!\{ m' \mid {}$}}
      \forall m''.\, \phase[m''] \,{\in}\, \{\Proposed, \Accepted\}$}
      \\
      \nonl \mbox{$\phantom{\mbox{{\bf forall} $\!\{ m' \mid \forall m''.\,$}}
      \!{\implies} \localts[m''] > \globalts[m']\}$}\\
      \nonl $\quad$\textbf{\em ordered by} $\globalts[m']$}{
      $\delivered[m'] \leftarrow \true$\;\label{alg:delivered-true}
      \Send $\DELIVER(m', \aballot, $\label{alg:send-deliver}
      \nonl{}$~~~~~~\localts[m'], \globalts[m'])$ \KwTo $g_0$; 
    }}

  \smallskip
  \smallskip
  \smallskip

  \SubAlgo{\WhenReceived $\DELIVER(m, b, \lts, \gts)$\label{alg:receive-deliver}}{%
    \precond $\status \in \{\FOLLOWER, \LEADER\} \wedge {}$\label{alg:check-deliver}\\ 
    \nonl $\phantom{\text{\precond}} \aballot = b \wedge \lastgts < \gts$\;
    $\phase[m] \leftarrow \Committed$\;\label{alg:set-committed-phase-recovery}
    $\localts[m] \leftarrow \lts$\;\label{alg:set-committed-lts}
    $\globalts[m] \leftarrow \gts$\;
    $\clock \leftarrow \max\{\clock, \ttime(\gts)\}$\;
    $\lastgts \leftarrow \gts$\;\label{alg:bump-last-gts}
    {\tt deliver}$(m)$;\label{alg:deliv-follower}
  }

  \smallskip
  \smallskip
  \smallskip

  \SubAlgo{\Fun ${\tt retry}(m)$\label{alg:retry}}{
    \precond $\phase[m] \in \{\Proposed, \Accepted\}$\;
    \mbox{\Send $\MULTICAST(m)$ \KwTo $\{\currleaders[g] \mid g \in \dest(m)\}$;}
  }
\end{algorithm}
}
\end{minipage}
&
\begin{minipage}{9cm}
\scalebox{0.98}{%
\removelatexerror
\begin{algorithm}[H]
  \SubAlgo{\Fun ${\tt recover}()$\label{alg:new-ballot} }{
    \Send $\NEWLEADER(\mbox{any ballot of the form $(\_, p_i)$}$\\
    \nonl $\phantom{\mbox{\bf send } \NEWLEADER(}\mbox{higher than $\ballot$})$ \KwTo $g_0$;
  }

  \smallskip
  \smallskip
  \smallskip

  \SubAlgo{\WhenReceived $\NEWLEADER(b)$ \KwFrom $p_j$\label{alg:newleader}}{
    \precond $b > \ballot$\;\label{alg:check-newleader}
    $\status \leftarrow \RECOVERING$\;
    $\ballot \leftarrow b$\;\label{alg:accept-new-ballot}
    \Send $\NEWLEADERACK(\ballot, \aballot, \clock, $\\
    \nonl $~~~~~~\,\phase, \localts, \globalts)$ \KwTo $p_j$;
  }

  \smallskip
  \smallskip
  \smallskip

  \SubAlgo{\WhenReceived
    $\NEWLEADERACK(b, \vaballot(p_j), $\label{alg:newleaderack}\\
    \nonl \quad$\vclock(p_j), \vphase(p_j), \vlocalts(p_j), \vglobalts(p_j))$\\
    \nonl $\quad$\FromQuorum $p_j  \in g_0$}{
    \precond $\status = \RECOVERING \wedge \ballot = b$\;\label{alg:newleaderack-check}
    reinitialise $\phase, \localts, \globalts$\;
    {\bf var} $J \leftarrow \mbox{the set of $j$ with maximal $\vaballot(p_j)$}$\;\label{alg:def-j}
    \ForAll{$m$}{
      \uIf{$\exists j.\, \vphase(p_j)[m] = \Committed$\label{alg:check-committed}}{
        $\phase[m] \leftarrow \Committed$\;\label{alg:recovery-phase-committed}
        $\localts[m] \leftarrow \vlocalts(p_j)[m]$\;\label{alg:recovery-lts-committed}
        $\globalts[m] \leftarrow \vglobalts(p_j)[m]$\; \label{alg:recovery-gts-committed} 
      }
      \ElseIf{$\exists j \in J.\, phase(p_j)[m] = \Accepted$\label{alg:check-accepted}}{
        $\phase[m] \leftarrow \Accepted$\;
        $\localts[m] \leftarrow \vlocalts(p_j)[m]$\;\label{alg:recovery-lts-accepted}
      }}
    $\clock \leftarrow \max\{\vclock(p_j)\}$\;\label{alg:clock-max-recovery}
    $\aballot = b$\;\label{alg:set-aballot-leader}
    \mbox{\Send $\NEWSTATE(b, \clock, \phase, \localts, \globalts)$}
    \KwTo $g_0 \setminus \{p_i\}$\; \label{alg:send-newview}
  }

  \smallskip
  \smallskip
  \smallskip

  \SubAlgo{\WhenReceived $\NEWSTATE(b, \vclock, \vphase, \vlocalts,
    \vglobalts)$ \KwFrom $p_j$\label{alg:trans-newleadersync}}{
    \precond $\status = \RECOVERING \wedge \ballot = b$\;\label{alg:check-newleadersync}
    $\status \leftarrow \FOLLOWER$\;
    $\aballot \leftarrow b$\;
    $\clock \leftarrow \vclock$;
    $\phase \leftarrow \vphase$;
    $\localts \leftarrow \vlocalts$;
    $\globalts \leftarrow \vglobalts$\;\label{alg:reinit-state}
    \Send $\NEWSTATEACK(b)$ \KwTo $p_j$\;
  }

  \smallskip
  \smallskip
  \smallskip

  \SubAlgo{\WhenReceived $\NEWSTATEACK(b)$\label{alg:receive-newstateack}\\
    \nonl $\quad${\bf from a set of processes that}\\
    \nonl $\quad${\bf together with $p_i$ form a quorum in $g_0$}}{
    \If{$\status = \RECOVERING \wedge \ballot = b$}{
     $\status \leftarrow \LEADER$\;
     \ForAll{$\!\{m' \mid \phase[m'] = \Committed \wedge  {}$\\ \label{alg:deliv-check2}
      \nonl \mbox{$\phantom{\mbox{{\bf forall} $\!\{ m' \mid {}$}} 
        \forall m''.\, \phase[m''] = \Accepted$}\\
        \nonl \mbox{$\phantom{\mbox{{\bf forall} $\!\{ m' \mid {}$}} \!{\implies}
          \localts[m''] > \globalts[m'])\}$}\\ 
      \nonl $\quad$\textbf{\em ordered by} $\globalts[m']$}{
       $\delivered[m'] = \true$\;
       \Send $\DELIVER(m', \aballot, $\label{alg:send-deliver-recovery} 
       \nonl{}$~~~~~~\localts[m'], \globalts[m'])$ \KwTo $g_0$; 
     }}}

  \smallskip
\end{algorithm}
}
\end{minipage}
\end{tabular}
\\
\caption{White-box multicast protocol at a process $p_i \in g_0$.}
\label{fig:protocol}
\end{figure*}


\paragraph{Preliminaries.}
Every process in a group is either the {\em leader} of the group or a {\em
  follower}. If the leader fails, one of the followers takes over. A major
design decision we take in our protocol is to use the {\em passive replication}
approach~\cite{zab,vr}: {\em only the leader computes the timestamps and decides
  when to deliver an application message}. Followers are passive: they merely
store the leader’s decisions, so that upon the leader failure a new leader could
recover the information necessary to continue multicast. A process maintains the
same variables as in Skeen's protocol (Figure~\ref{fig:skeen}) and a few
additional ones. A $\status$ variable records whether the process is a
$\LEADER$, a $\FOLLOWER$ or is in a special $\RECOVERING$ state used during
leader changes. A period of time when a particular process $p_i$ acts as a
leader is denoted using a {\em ballot} $(n, p_i)$---a pair of an integer $n$ and
the process identifier $p_i$. Ballots are ordered lexicographically using an
arbitrary total order on processes, with a special ballot $\bot$ being the
minimal ballot. For a ballot $b = (n, p_i)$ we let $\leader(b) = p_i$. At any
given time, a process participates in a single ballot, which is stored in a
variable $\aballot$ and never decreases. During leader changes we also use an
additional ballot variable $\ballot$.

\paragraph{Normal operation.} To multicast an application message $m$, a process
sends it in a $\MULTICAST$ message to the current leader of every group
$g\in \dest(m)$ (line~\ref{alg:client}), which is determined using a mapping
$\currleaders$. This mapping need only give a guess as to the identity of the
current leaders. If the guess is wrong, the multicasting process can always
send the message to all the processes in a given group to find out who its
leader is (omitted from the pseudocode).


A process $p_i$ handles the message only when it is indeed the leader of its
group $g_0$ (line~\ref{alg:leader-origin}). When the leader receives $m$ for the
first time (line~\ref{alg:lts-exists}), it performs the same actions as 
in Skeen's
protocol (lines~\ref{sk:clock-inc}-\ref{sk:set-phase-proposed} in
Figure~\ref{fig:skeen}): it increments the clock, computes the local timestamp,
and sets the phase of $m$ to $\Proposed$.

Like in Skeen's protocol, the leader's next goal is to communicate its local
timestamp proposal to the leaders of the other destination groups of $m$.
A key idea used to achieve fault-tolerance and reduced latency in our protocol
is not to send local timestamps to the leaders directly, but {\em route them
  through a quorum of processes in each destination group, to ensure their
  durability.} Namely, the leader sends an $\ACCEPT$ message including its
ballot and the computed local timestamp to all processes in $\dest(m)$
(including itself, for uniformity, line~\ref{alg:send-accept}); this message is
analogous to the ``2a'' message of Paxos. As we explain in the following, due to
failures the leader may receive the same $\MULTICAST(m)$ message twice. In this
case the leader resends the $\ACCEPT$ message with the locally stored data
for $m$. This ensures Invariant~\ref{inv:unique-lts}: in a given ballot, a
message can be assigned at most one local timestamp.

A process that is a destination of $m$ acts once it receives an $\ACCEPT$
message for $m$ from the leader of each of the destination groups
$g \in \dest(m)$ (line~\ref{alg:receive-accept}). The message carries the local
timestamp proposal $\Lts(g)$ and the ballot $\Bal(g)$ of the leader making the
proposal. The process checks that it participates in the ballot $\Bal(g_0)$ of
the leader of its group $g_0$ it received the message from. Then the process
advances the phase of the message $m$ to $\Accepted$, stores its local timestamp
in the $\localts$ array (line~\ref{alg:store-lts}) and ensures its clock is no
lower than the global timestamp obtained by taking the maximum of the local
timestamps $\Lts(g)$ of $m$ (line~\ref{alg:clock-max}). Thus,
lines~\ref{alg:store-lts} and~\ref{alg:clock-max} in our protocol can be viewed
as replicating lines~\ref{sk:origin} and~\ref{sk:clock-max} of Skeen's protocol
(Figure~\ref{fig:skeen}) throughout the process group. The process acknowledges
the acceptance of the local timestamps by sending an $\ACCEPTACK$ message to the
leaders who made the proposals, tagged with the vector of ballots $\Bal$ in
which these proposals were made at the destination groups; this message is
analogous to the ``2b'' message of Paxos.

A leader who made a local timestamp proposal for $m$ waits until it receives a
quorum of $\ACCEPTACK$ messages for $m$ with matching ballot vectors from each
of the destination groups $\dest(m)$ (line~\ref{alg:receive-acceptack});
Invariant~\ref{inv:unique-lts} ensures that the different $\ACCEPTACK$ messages
correspond to the same set of local timestamp proposals. At this point the
leader considers that all local timestamps for $m$ are agreed, and thus it
advances the phase of $m$ to $\Committed$, computes its final global timestamp
as the maximum of the local timestamps and stores it in the $\globalts$
array. The leader then tries to deliver one or more committed messages like in
Skeen's protocol, in the order of their global timestamps
(line~\ref{alg:deliv-check1}, corresponding to line~\ref{sk:deliv-check1} in
Figure~\ref{fig:skeen}). To this end, it sends the data about each message $m'$
to deliver in a $\DELIVER$ message to all the members of its group.

Since our communication channels are FIFO, during failure-free execution a
process receives $\DELIVER$ messages in the order the leader of its group sends
them. Upon receiving such a message, the process stores the enclosed information
and delivers the corresponding application message
(line~\ref{alg:receive-deliver}). As we explain in the following, when failures
occur, a process may receive duplicate $\DELIVER$ messages. To handle this, each
process maintains the highest global timestamp of an application message it has
delivered in a variable $\lastgts$ and ignores $\DELIVER$ messages carrying
lower global timestamps.

\paragraph{Discussion of normal operation.}
As we mentioned earlier, our optimised protocol can be viewed as weaving
together the steps from Skeen's protocol and Paxos. In particular, when
multicasting a local application message $m$ with $\dest(m) = \{g_0\}$, the
protocol exactly follows the flow of Paxos: the leader of $g_0$ sends a proposal
to all processes in $g_0$ ($\ACCEPT$) and waits for a quorum of acknowledgements
($\ACCEPTACK$), whereupon it delivers $m$ ($\DELIVER$). Like in Paxos, when a
process receives the $\ACCEPT$ message from the leader
(line~\ref{alg:receive-accept}), the process checks that it participates in the
ballot the leader is in (line~\ref{alg:check-propose}), thus ensuring that it
only stores local timestamps (line~\ref{alg:store-lts}) issued by the leader it
supports.

For a global application message, the flow of the protocol is also similar to
the one of Paxos, but performed between multiple leaders on the one hand and
multiple groups of followers on the other. However, note that a process does not
perform any checks on ballots in $\ACCEPT$ messages received from remote groups
(line~\ref{alg:receive-accept}); these ballots are only used in
line~\ref{alg:receive-acceptack} to ensure that different $\ACCEPTACK$ messages
correspond to the same set of local timestamp proposals. Hence, the $\ACCEPT$
messages may well come from old leaders of remote groups that have since been
deposed and whose local timestamp proposals will be rejected by their
groups. The update to the clock at line~\ref{alg:clock-max} may thus be
performed based on such invalid local timestamps. A key insight used in our
protocol is that this situation does not violate correctness. The $\clock$
variables at processes of the same group are used to simulate the $\clock$
variable of a reliable process in Skeen's protocol: as we explain in the
following, if the group leader fails, a new leader recovers the clock value from
the clocks at followers. But as we noted in \S\ref{sec:skeen}, {\em the $\clock$
  variable in Skeen's protocol can always be safely increased.}

Hence, the Paxos-like $\ACCEPT$ and $\ACCEPTACK$ messages in our protocol can be
viewed as replicating in one go both the local timestamp assignment
(line~\ref{sk:origin} in Figure~\ref{fig:skeen}) and the clock increase
(line~\ref{sk:clock-max} in Figure~\ref{fig:skeen}), with the latter done
speculatively, before the local timestamps are agreed. Once a leader receives a
quorum of $\ACCEPTACK$ messages from each of the destination groups
(line~\ref{alg:receive-acceptack} of our protocol), it knows that the clocks at
the processes in these quorums have already been advanced to be no lower than
the corresponding global timestamp. The leader can thus avoid a round trip to
replicate the clock update, required in the naive fault-tolerant version of
Skeen's protocol we presented earlier. The leader then replicates the global
timestamps off the critical path, in $\DELIVER$ messages, by exploiting the fact
that global timestamps are uniquely determined by local timestamps.


\begin{figure}[t]
\begin{enumerate}
\item \label{inv:unique-lts} For any two messages sent of the form
  $\ACCEPT(m, g, b, \lts_1)$ and $\ACCEPT(m, g, b, \lts_2)$, we must have
  $\lts_1 = \lts_2$.

\item\label{inv:main} Assume that at some point a quorum of processes in $g_0$
  have received the set of messages
\begin{equation}\label{inv-main-accept}
\{\ACCEPT(m, g, \Bal(g), \Lts(g)) \mid g \in \dest(m)\}
\end{equation}
and responded to them with
\begin{equation}\label{inv-main-accept-ack}
\ACCEPTACK(m, g_0, \Bal).
\end{equation}
Whenever at a process in $g_0$ we have $\aballot > \Bal(g_0)$, we also have:
\begin{enumerate}
\item\label{main-phase}
$\phase[m] \in \{\Accepted, \Committed\}$;
\item\label{main-lts}
$\localts[m] = \Lts(g_0)$;
\item\label{main-clock}
$\clock \ge \ttime(\max\{\Lts(g) \mid g \in \dest(m)\})$;
\end{enumerate}

\item\label{inv:consensus} 
  \begin{enumerate}
  \item For any messages $\DELIVER(m, \_, \lts_1, \_)$ and
    $\DELIVER(m, \_, \lts_2, \_)$ sent to processes in the same group, we have
    $\lts_1 = \lts_2$.

  \item For any messages $\DELIVER(m, \_, \_, \gts_1)$ and
    $\DELIVER(m, \_, \_, \gts_2)$ sent to any groups, we have $\gts_1 = \gts_2$.
\end{enumerate}

\item\label{inv:unique-gts}
  For any $\DELIVER(m_1, \_, \_, \gts_1)$ and $\DELIVER(m_2, \_, \_, \gts_2)$
  messages sent, if $m_1 \not= m_2$, then $\gts_1\not=\gts_2$.

\item\label{inv:main-msg} Assume that at some point a quorum of processes in
  $g_0$ have received the set of messages~(\ref{inv-main-accept}) and responded
  to them with~(\ref{inv-main-accept-ack}) and that this quorum includes
  $\leader(\Bal(g_0))$. Let $\gts = \max\{\Lts(g) \mid g \in \dest(m)\}$ and let
  $\vlocalts_0$ be the projection of $\localts$ when $\leader(\Bal(g_0))$ sent
  its $\ACCEPTACK$ to messages $m'$ such that
  $\phase[m'] \not= \Start \wedge \localts[m'] < \gts$. Whenever at a process in
  $g_0$ we have $\aballot > \Bal(g_0)$, we also have:
\begin{multline}\label{main-msg}
  \forall m'.\, \phase[m'] \not= \Start \wedge \localts[m'] < \gts \\
  {} \implies \localts[m'] = \vlocalts_0[m'].
\end{multline}

\item\label{inv:main-live} Starting from some time $t \ge \GST$, for every group
  $g\in \G$ there exists a quorum $Q\subseteq g$ of correct processes, 
  and $p_i\in Q$ such that all
  members of $Q$ permanently follow $p_i$ as their leader,
  and all correct processes in $\Proc$ have their $\currleaders[g] = p_i$.

\end{enumerate}
\setcounter{saveenum}{\value{enumi}}
\caption{Key invariants of the white-box multicast protocol.}
\label{fig:inv}
\end{figure}

\paragraph{Key invariants.} 
We now describe the key invariants of the protocol used to prove its
correctness, which also motivate the design of recovery from leader failures.
Invariant~\ref{inv:main} ensures that, if a quorum of processes in a group $g_0$
accepted the same set of local timestamp proposals $\Lts$ for an application
message $m$, then the message $m$ and its local timestamp $\Lts(g_0)$ at $g_0$
will persist in all ballots higher than the ballot $\Bal(g_0)$ at which $g_0$
accepted them (a, b); furthermore, the clock values at these ballots will be no
lower than the global timestamp computed from the local timestamp proposals
$\Lts$ for $m$ (c). Lines~\ref{alg:set-accept-commit},~\ref{alg:store-lts}
and~\ref{alg:clock-max} in our protocol contribute to preserving the clauses
(a), (b) and (c) of the invariant, respectively. Since
Invariant~\ref{inv:main}(a, b) ensures that local timestamps accepted by a
quorum persist across leader changes, we then get
Invariant~\ref{inv:consensus}(a), ensuring that each group agrees on the local
timestamp of a given application message. Since the global timestamp for an
application message is computed as the maximum of local timestamps accepted by
quorums in each destination group, from Invariant~\ref{inv:consensus}(a) we get
Invariant~\ref{inv:consensus}(b), ensuring that the system agrees on the global
timestamp of each message. Finally, similarly to how it was done for Skeen's
protocol, we can show Invariant~\ref{inv:unique-gts}, ensuring that global
timestamps are unique for each message.

Finally, Invariant~\ref{inv:main-msg} ensures that application messages are
delivered in the order of their global timestamps, despite leader
changes. Similarly to Invariant~\ref{inv:main}, this invariant assumes that a
quorum of processes in a group $g_0 \in \dest(m)$, including its leader
$\leader(\Bal(g_0))$, have accepted the same set of local timestamp proposals
$\Lts$ for $m$, yielding a global timestamp $\gts$.  The invariant ensures that,
in any future ballot of group $g_0$, a process may not have messages with local
timestamps less than $\gts$ that the leader $\leader(\Bal(g_0))$ did not know
about when it accepted the local timestamp for $m$. Given this invariant and the
check on local timestamps the leader performs before delivering an application
message (line~\ref{alg:deliv-check1}), if a leader of a group $g_0$ delivers a
message $m$ with a global timestamp $\gts$, then it can be sure that no message
it is not aware of will get a local timestamp lower than $\gts$ in future
ballots, and thus no message will get a lower global timestamp.

Invariant~\ref{inv:main-msg} is proved using Invariant~\ref{inv:main}(c): under
the assumptions of the former invariant, the latter one ensures that the clock
of any leader of a future ballot will be no lower than $\gts$. Then any new
application message this leader receives will get a local timestamp at $g_0$
higher than $\gts$.

\paragraph{Leader recovery.} 
We assume that each group $g\in \G$ is equipped with a {\em leader selection
service (LSS)}, which is responsible for nominating  a single member of $g$ as
a potential new leader. The LSS implementation exploits the knowledge
of the upper bound on the  failure-free message propagation delay $\delta$ to
guarantee that eventually the same correct member of $g$ is permanently
suggested by LSS as a leader of $g$ to all its correct members. 
Examples of LSS implementations
satisfying this property can be found
in~\cite{CT96,Larrea04,Larrea00,Aguilera01}.

A leader recovery procedure for a group $g$ is activated whenever LSS nominates
a new process as a leader candidate, or the current leader fails
to collect a quorum of responses to one of its messages.
The main goal of the procedure is to
preserve Invariants~\ref{inv:main} and~\ref{inv:main-msg}. Ensuring the latter
is particularly subtle: for this, {\em before the new leader starts multicast,
it   must bring a quorum of followers in sync with its state} (this is similar
to~\cite{vr,zab}). Hence, a new leader is elected in two stages. First,
processes vote to join the ballot of a prospective leader, which they record
in a variable $\ballot$; like $\aballot$, this variable can only increase.
Second, processes receive and acknowledge an initial state from the new leader
and set $\aballot$ to $\ballot$. The leader only resumes normal operation after
it gets a quorum of such acknowledgements. Note that we thus always have
$\aballot \le \ballot$.

In more detail, a process $p_i$ initiates the recovery procedure
by invoking 
the {\tt recover} function (line~\ref{alg:new-ballot}), which attempts to 
establish a new ballot with $p_i$ as its leader.
The process picks a ballot that it leads and
higher than the last ballot it joined, and sends the ballot in a $\NEWLEADER$ message to
the group members (including itself); this message asks the group members to
support the process as the new leader and is analogous to the ``1a'' message in
Paxos. When a process receives a $\NEWLEADER(b)$ message
(line~\ref{alg:newleader}), it first checks that the proposed ballot $b$ is
higher than the last ballot it joined. In this case it sets $\ballot$ to $b$ and
changes its status to $\RECOVERING$, which causes it to stop normal message
processing.
The process then replies to the new leader with a
$\NEWLEADERACK$ message containing all components of its state; this message
serves as a vote for the new leader and is analogous to the ``1b'' message of
Paxos.

The new leader waits until it receives $\NEWLEADERACK$ messages from a quorum of
group members (line~\ref{alg:newleaderack}). Based on the states reported in
them, it computes a new state from which to resume multicast according to the
following rules. First, if an application message $m$ is $\Committed$ at some
process, then the leader marks it as $\Committed$ and copies its local and
global timestamps (line~\ref{alg:check-committed}). If a message $m$ is not
$\Committed$ at any process, then, like in Paxos, the leader looks at the states
of processes that reported the maximal $\aballot$ (line~\ref{alg:def-j}): if a
message $m$ is $\Accepted$ at such a process, then the leader marks it as
$\Accepted$ and copies its local timestamp (line~\ref{alg:check-accepted}). Like
for Paxos, we can show that these rules preserve Invariant~\ref{inv:main}(a,
b). Finally, the leader sets $\clock$ to the maximum of the clock values
reported by processes, to preserve Invariant~\ref{inv:main}(c), and sets
$\aballot$ to the new ballot.

The new leader next ensures that at least a quorum of processes in its group are
in sync with its new state. To this end, it sends a $\NEWSTATE$ message with the
new state to the other group members (line~\ref{alg:send-newview}). Upon
receiving this message (line~\ref{alg:trans-newleadersync}), a process
overwrites its state with the one provided, changes its status to $\FOLLOWER$,
and sets $\aballot$ to $b$, thereby recording the fact that it has synchronised
with the leader of $b$. The process then replies to the new leader with a
message $\NEWSTATEACK(b)$ confirming this.


The new leader waits until it receives $\NEWSTATEACK$ from a set of processes
that together with it form a quorum (line~\ref{alg:receive-newstateack}). The
leader may have application messages ready to be delivered that some of the
followers have not delivered yet. In fact, different followers may have
delivered different sequences of application messages, because the previous
leader may have crashed in between sending $\DELIVER$ messages to different
followers. To deal with this, the leader delivers all committed messages it
can, starting from the beginning. This does not violate correctness since, as we
explained earlier, followers check for duplicate $\DELIVER$ messages using the
$\lastgts$ variable. At the end, the new leader sets $\status$ to $\LEADER$,
which allows it to start normal operation.

If at any point in the execution of the protocol, the current leader 
(or a leader candidate) time-outs on acquiring a quorum of responses 
to one of its messages,
it falls back to leader recovery with a higher ballot by invoking the
\texttt{recover} function. This guarantees that Invariant~\ref{inv:main-live} in
Figure~\ref{fig:inv} holds, which ensures that a stable leader will eventually
be established.

\paragraph{Discussion of leader recovery.} 
We now highlight some of the subtleties of the recovery procedure. First, note
that upon a leader change, the value of the clock at leaders may actually
decrease. For example, assume $p_i \in g_0$ is a leader who issued a local
timestamp $(t, g_0)$ for $m$ and thus set $\clock = t$. If $p_i$ fails before a
quorum of processes in $g_0$ accepts $m$, the new leader may derive its initial
state from a quorum of processes that did not see $m$ and end up with a clock
value lower than $t$. This does not violate correctness: to ensure that messages
are delivered in the order of their timestamps we only need to ensure that the
clock does not fall below the global timestamp of a message accepted by a
quorum, as stated by Invariant~\ref{inv:main}(c).

We next illustrate why it is important for a leader to synchronise its state
with the followers before starting normal operation. Assume a process
$p_1 \in g_0$ is a leader of a ballot $b_1$ who has issued a local timestamp
$\lts$ for an application message $m$ and replicated it to some of its followers
in $g_0$. Assume further that before $p_1$ manages to reach a quorum, another
process $p_2 \in g_0$ becomes the leader at $b_2 > b_1$. To compute its initial
state, $p_2$ may query a quorum that does not contain any processes that saw $m$
and $\lts$, so that its initial state will exclude these. Assume that at a later
point $p_2$ commits and delivers a message $m'$ with a global timestamp
$\gts' > \lts$. Now imagine there is yet another leader change and a process
$p_3$ becomes a leader at a ballot $b_3 > b_2$. Since before $p_2$ delivered
$m'$, it got a quorum of followers to accept its initial state and set
$\aballot = b_2$, when $p_3$ queries a quorum to compute its initial state, it
is guaranteed to see at least one process with $\aballot = b_2$; this process
will report a state excluding $m$ and $\lts$. According to the rule used to
compute the initial state in line~\ref{alg:check-accepted}, $p_3$ will then
disregard any processes that accepted $m$ and $\lts$ at $b_1 < b_2$. This will
ensure Invariant~\ref{inv:main-msg}: the local timestamp $\lts$ for $m$, which
$p_2$ did not know about when it committed $m'$, will never be resurrected upon
recovery. Hence, the message $m$ will never be able to get a timestamp lower
than $\gts'$, and the decision by $p_2$ to deliver $m'$ will stay valid.


\paragraph{Message recovery.} In the above scenario message $m$ gets lost at the
group $g_0$ due to a leader failure. Even if other destination groups have
received it, its processing will not progress. To deal with this situation, the
multicasting process can just resend the $\MULTICAST(m)$ message. Then groups
that have not previously received $m$ will start processing it, and groups that
have already processed $m$ will just resend the corresponding protocol messages
(lines~\ref{alg:send-accept} and \ref{alg:send-acceptack}), which will unblock
the processing of $m$.

The processing of a message $m$ can also get stuck if the process submitting it
for multicast fails in between sending $\MULTICAST(m)$ messages to different
leaders (line~\ref{alg:send-multicast}), so that one group $g_1 \in \dest(m)$
receives $m$ and another group $g_2 \in \dest(m)$ does not receive it. This will
cause $m$ to get stuck in the $\Proposed$ phase at the leader of $g_1$, since
the group $g_2$ will never send a local timestamp proposal for $m$. The leader
of $g_1$ can again recover from this situation by resending the $\MULTICAST(m)$
message to all destination groups of $m$ (line~\ref{alg:retry}). The same
mechanism can be used to resume the processing of an accepted message after a
leader change.


\section{Correctness and Latency Analysis}
\label{sec:correctness}

\begin{theorem}\label{thm:correctness}
  The white-box protocol in Figure~\ref{fig:protocol} is a correct and genuine
  implementation of atomic multicast.
\end{theorem}
\smallskip
\smallskip
Due to space constraints, we defer the proof of the theorem
to~\tr{\ref{app:correctness}}{\ncorrectness}. The proofs of the Ordering,
Validity and Integrity properties rely on
Invariants~\ref{inv:unique-lts}-\ref{inv:main-msg} in Figure~\ref{fig:inv}. The
proof of Termination relies on Invariant~\ref{inv:main-live}, which implies 
\smallskip
\smallskip
\begin{lemma}
  Let $t$ be the time stipulated by Invariant~\ref{inv:main-live} 
  in Figure~\ref{fig:inv}. Then for all
  application messages $m$ and all groups $g\in \dest(m)$, if $p_i$ is the
  leader of $g$, and $m$ is either known to $p_i$ at $t$ or received at $p_i$
  after $t$ via $\MULTICAST(m)$, then $m$ is eventually committed at $p_i$.
\label{lem:all-commit}
\end{lemma}
\smallskip
\smallskip

In particular, the lemma implies that there exists a time $t' \ge t$ such that
after $t'$, the leaders of all groups in $\G$ do not have any uncommitted
messages that were multicast before $t$. The lemma and this consequence hold not
just for our white-box protocol, but also for other protocols we consider in
this paper, such as the naive fault-tolerant version of Skeen's protocol from
\S\ref{sec:optimised}. Using this fact, we now show how to establish the
collision-free and failure-free latencies of a Skeen-based protocol $\A$ by
analysing the delivery latency of a message multicast after $t'$.

We first consider the collision-free case. 
Let $D$ be the {\em commit latency} of $\A$, 
i.e., the maximum amount of time elapsing between the
events of multicasting a message $m$ after $t$ and $m$ being committed by the
leader of some group in $\dest(m)$. Consider
an application message $m$ that was multicast at time $t_1 > t'$ and let
$p_i$ be the leader of a group $g\in \dest(m)$. By Lemma~\ref{lem:all-commit},
there exists a time $t_2 > t_1$ at which $p_i$ commits
$m$. Suppose that $m$ is not
concurrent with any conflicting messages that were multicast by correct processes.
Consider an arbitrary message $m'$ known to $p_i$ at $t_2$. 
If $m'$ was multicast by a correct process after $t$, it must have been 
delivered, and therefore committed at $p_i$, before $t_1$.
Otherwise, $m'$ is multicast either before $t$ or by a faulty process.
Since failures stop after $\GST$, the latter implies that
$m'$ was multicast prior to $\GST \le t$. Thus, $m'$ must have 
been multicast before $t$ in both cases, which,
by the choice of $t'$, implies that $m'$ was committed at $p_i$ before $t_1 > t'$.
Thus, at $t_2$, $p_i$ does not have any uncommitted messages other than $m$, and
therefore can deliver $m$. Since $t_2 \le t_1 + D$, we have
\smallskip
\smallskip
\begin{theorem}\label{thm:cfl}
  The collision-free latency of a Skeen-based atomic multicast implementation
  $\A$ is equal to $\A$'s commit latency $D$.
\end{theorem}
\smallskip
\smallskip

We next give a method for computing the failure-free latency of a Skeen-based
protocol $\A$. Let $C$ be the {\em clock update latency} of $\A$, i.e., the
maximum amount of time elapsing between the events of multicasting a message $m$
after $t$ and advancing the clock past $\globalts[m]$ at the leader of some
group in $\dest(m)$. Consider
an application message $m$ that was multicast at time $t_1 > t'$ and let
$p_i$ be the leader of a group $g\in \dest(m)$.
By Lemma~\ref{lem:all-commit}, there exist times $t_c > t_1$ and 
$t_2 \ge t_c$ such that $p_i$ 
advances its clock past $\globalts[m]$ at $t_c$ and commits $m$ at $t_2$.
The delivery of message $m$ can be delayed past its commit time only by a
conflicting concurrent message multicast after $t$. 
Consider such a message $m'$ and let $t'_2$ be the time at which it commits at
$p_i$ (which exists by Lemma~\ref{lem:all-commit}). 
If $p_i$ receives $\MULTICAST(m')$ after
$t_c$, then at the time $m$ is committed, $\localts[m'] > \globalts[m]$, and
therefore, $p_i$ does not need to wait until $m'$ is committed to deliver $m$.
Suppose now that $p_i$ receives $\MULTICAST(m')$ before $t_c$, and 
let $t_1'$ be the time when $\multicast(m')$ occurs. If $t_1' < t$, 
then by the choice of $t'$, $m'$ is committed at $p_i$ at time $t_2 > t'$, 
and therefore, will not be obstructing the delivery of $m$. Otherwise, 
$t_2' \le t_1' + D$, and therefore, $t_2'$ is maximised if $t_1'$ is arbitrarily close to $t_c$.
We thus have $t_2' \le t_c + D \le t_1 + C + D$.
Hence, at the latest $m$ is delivered at $p_i$ at $\max\{t_2, t_1 + C +
D\} \le \max\{t_1 + D, t_1 + C + D\} = t_1 + C + D$. 
Then Theorem~\ref{thm:cfl} implies
\smallskip
\smallskip
\begin{theorem}
\label{thm:ffl}
  The failure-free latency \textit{FFL} of a Skeen-based atomic multicast
  implementation $\A$ is $\textit{FFL} = C + \textit{CFL}$, where $C$ is the
  clock update latency of $\A$, and $\textit{CFL}$ is its collision-free
  latency.
\end{theorem}
\smallskip
\smallskip

The commit latency of our white-box protocol is $3\delta$, corresponding to the
sequence of messages $\MULTICAST$, $\ACCEPT$, $\ACCEPTACK$. In contrast, its
clock update latency is $2\delta$, corresponding to the messages $\MULTICAST$
and $\ACCEPT$ (see line~\ref{alg:clock-max} of Figure~\ref
{fig:protocol}). Hence, Theorems~\ref{thm:cfl} and \ref{thm:ffl} imply
\smallskip \smallskip
\begin{theorem}
The collision-free latency of the white-box protocol 
in Figure~\ref{fig:protocol} is $3\delta$, and its 
failure-free latency is $5\delta$.
\end{theorem}
\smallskip
\smallskip
Since in our protocol followers deliver an application message only
after receiving a $\DELIVER$ message from their leader, the maximum time to
deliver a message at followers is $4\delta$ in a collision-free run and $6\delta$ in a
failure-free one.

In contrast to the white-box protocol, the naive fault-tolerant version of
Skeen's protocol from \S\ref{sec:optimised} has the commit latency of $6\delta$,
which by Theorem~\ref{thm:cfl} equals its collision-free latency. In this
protocol a leader advances its clock past a message's global timestamp only
after completing the corresponding consensus call, resulting in the clock update
latency of $6\delta$. Hence, by Theorem~\ref{thm:ffl}, the failure-free latency
of fault-tolerant Skeen's protocol is $12\delta$.


\section{Experimental Evaluation}
\label{sec:exp}

We have implemented our multicast protocol in C using the {\tt libevent} library
for communication~\cite{libevent}. Our implementation is available
at~\cite{implementation}. In addition to the protocol described in
\S\ref{sec:optimised}, the implementation includes a mechanism to garbage
collect delivered messages. In this section we experimentally compare our
protocol with the naive fault-tolerant version of Skeen's we described in
\S\ref{sec:optimised}~\cite{raynal} and a state-of-the-art {\em FastCast}
protocol by Coelho et al.~\cite{fastcast}. We use open-source implementations of
these protocols by Coelho et al.~\cite{code}, also implemented in C and using
{\tt libevent}.

\begin{figure*}[t]
\includegraphics[width=\textwidth]{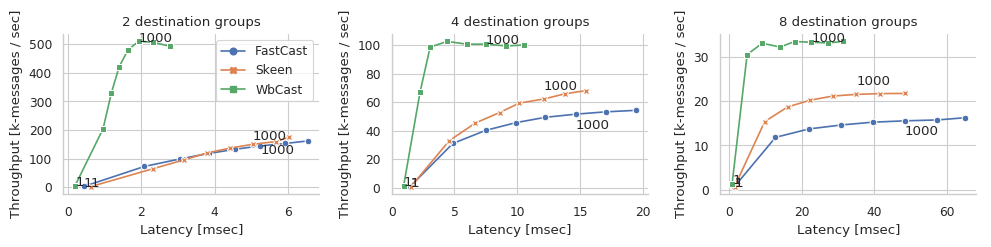}
\caption{Performance of multicast protocols in LAN with increasing numbers of
  clients: FastCast, fault-tolerant Skeen and our protocol (WbCast). In each
  experiment clients multicast messages to a fixed number of groups. For
  reference, we mark the points corresponding to 1000 clients.}
\label{fig:graph-lan}
\end{figure*}

\begin{figure*}[t]
\includegraphics[width=\textwidth]{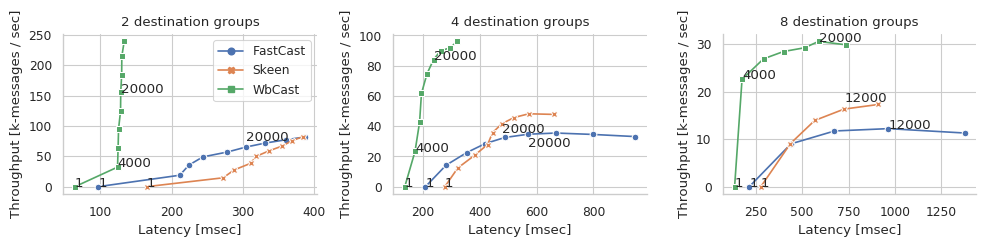}
\caption{Performance of multicast protocols in WAN with increasing numbers of
  clients: FastCast, fault-tolerant Skeen and our protocol (WbCast). In each
  experiment clients multicast messages to a fixed number of groups. For
  reference, we mark the points corresponding to certain numbers of clients.}
\label{fig:graph-wan}
\end{figure*}

\paragraph{FastCast.} We first briefly review the FastCast protocol. This
protocol optimises fault-tolerant Skeen's by using speculative execution, while
still using consensus as a black box. Like in Skeen's protocol, upon receiving
an application message, the Paxos leader of a group issues a local timestamp
based on its local clock and invokes consensus to persist it. However, the
leader also immediately sends the local timestamp to the leaders of the other
destination groups, without waiting for consensus to finish. The leaders then
speculatively act on these timestamps like in Skeen's, computing the global
timestamp as their maximum, advancing their clocks in line with it and invoking
consensus to persist these actions. Once the consensus on the local timestamps
is reached, the leaders exchange messages confirming this. By the time a leader
receives these messages, it may have already done all of the work necessary to
act on the local timestamps, and can commit the corresponding application
message at once. Using the method in \S\ref{sec:correctness}, we can show that
FastCast's collision-free and failure-free latencies are $4\delta$ and
$8\delta$, respectively.

\paragraph{Local-area network.} We first benchmark the protocols in a local-area
network (LAN) using the CloudLab infrastructure~\cite{cloudlab}. We consider 10
groups, each with 3 replicas, residing on 30 machines. A varying numbers of
client processes residing on 10 separate machines initiate multicasts of 20-byte
messages in a closed loop. We use machines with 10-core Xeon E5-2640 processors
and 64GB of memory, connected by 2GB network links with around 0.1ms round-trip
time.

We follow the evaluation methodology similar to the one previously used to
benchmark FastCast~\cite{fastcast}. In Figure~\ref{fig:graph-lan} we show the
average latency and throughput in 3-minute long runs as a function of the number
of clients and the number of destination groups these clients multicast
to. Additional graphs for other numbers of destination groups are provided
in~\tr{\ref{app:exp}}{\nexp}. All protocols we consider are CPU-bound in this
experiment, reaching 100\% utilization when saturated.


As is evident from Figure~\ref{fig:graph-lan}, our protocol consistently
outperforms FastCast and Skeen both in latency and in throughput. For example,
at 1000 clients our protocol outperforms FastCast by 1.2-3.5x, depending on the
number of destination groups, and by 2.15x on average. Note that in LAN,
FastCast generally performs slightly worse than Skeen. This is consistent with
the results in~\cite{fastcast} and is due to the overhead of introduced by its
parallel execution paths: this protocol is more suited for a wide-area network.

\paragraph{Wide-area network.} We next benchmark the protocols in a wide-area
network (WAN). We again consider 10 groups, which are replicated across 3 data
centres on the Google Cloud Platform. Each group has a replica in each data
centre, so that a single data centre contains a complete copy of the data
managed by the system. This setting is typical for modern wide-area
deployments~\cite{spanner}. The data centres are Oregon (R1), North Virginia
(R2) and England (R3), and average round-trip times between them are 60ms
(R1$\leftrightarrow$R2), 75ms (R2$\leftrightarrow$R3) and 130ms
(R1$\leftrightarrow$R3). We use 30 machines with 2 vCPUs and 7.5GB of memory for
multicast group members, and 3 machines per datacentre with 8 vCPUs and 30GB of
memory to generate client load.

In Figure~\ref{fig:graph-lan} we show the performance of all protocols in this
environment (additional graphs are provided in~\tr{\ref{app:exp}}{\nexp}). Our
protocol again outperforms both FastCast and Skeen. For example, at 8000 clients
it outperforms FastCast on both latency and throughput by 1.1-3.1x, depending on
the number of destination groups, and by 2x on average.

\paragraph{Recovery.} We have also benchmarked the performance of recovery in
the above WAN environment. In this experiment 6000 client threads multicast
messages to subsets of 4 out of 10 groups, and the leader of one of the groups
crashes. It takes 6 sec for the affected group to recover: 2.5 sec for a
new leader to get established (i.e., switch to the $\LEADER$ state), and 3.5
sec for the new leader to clear out the messages whose processing was
interrupted by the crash. We give a detailed graph of the performance in the
experiment in~\tr{\ref{app:exp}}{\nexp}.


\section{Related Work}
\label{sec:related}

Genuine atomic multicast is often implemented using a fault-tolerant version of
Skeen's protocol~\cite{raynal,multicast-cost}, which has the collision-free
latency of $6\delta$.  Early alternatives had asymptotically worse time
complexity, e.g., proportional to the number of destination
groups~\cite{delporte}. As this is unsatisfactory, researchers have been looking
for protocols with lower latency. Rodrigues et al.~\cite{scalcast} proposed a
protocol that has the collision-free latency of $5\delta$. More recently, Coelho
et al.~\cite{fastcast} proposed the FastCast protocol that further lowers it to
$4\delta$, which we discussed in detail in \S\ref{sec:exp}. In comparison to
this protocol, ours avoids using separate consensus calls to replicate a local
timestamp and to advance the clock above a global timestamp, resulting in
collision-free latency of $3\delta$. It also boasts a lower failure-free latency
of just $5\delta$, thus reducing the 2x latency degradation caused by concurrent
messages in existing atomic multicast implementations.

Our experimental results demonstrate that minimising latency is not only
of theoretical interest, but enables superior performance in practice. The above
protocols also used consensus as a black-box, whereas take a different approach,
unpacking Paxos and weaving it together with Skeen's protocol.

In this paper we assumed that each group has enough correct processes to
function normally. Researchers have also investigated atomic multicast protocols
that can operate when a whole group crashes~\cite{crashing-groups}. We also
assumed that process failures are crash-stop, rather than
Byzantine~\cite{byz-multicast}. We leave handling these more challenging cases
for future work.

Another primitive whose fault-tolerance presents similar challenges to atomic
multicast is {\em atomic commit}, which allows several process groups to reach a
decision on whether a database transaction should be committed or aborted. A
naive fault-tolerant solution to this problem layers the classical two-phase
commit protocol over Paxos~\cite{spanner}. There have been several alternative
proposals that reduce the latency by developing a single coherent protocol, in
the spirit of this work~\cite{uw-inconsistent,mdcc,commit}. In comparison to
these proposals, we handle the more challenging problem of atomic multicast,
where process groups need to agree on a total ordering of application messages
rather than on a binary per-transaction decision. This required us to develop
new techniques for replicating operations on logical clocks in a
latency-conscious way.



\paragraph{Acknowledgements.}

We thank our shepherd, Jos\'e Orlando Pereira, as well as Manuel Bravo, Thanh
Hai Tran and Pierre Sutra for helpful comments and discussions. We also thank
Paulo Coelho and Fernando Pedone for discussions about their FastCast
protocol. Alexey Gotsman was supported by an ERC grant RACCOON.

{\footnotesize
\bibliographystyle{abbrv}
\bibliography{biblio}
}

\iflong
\appendices
\clearpage
\onecolumn
\section{Proof of Correctness}
\label{app:correctness}

\smallskip
\smallskip

Our proof uses the following invariants of the protocol in addition to those in
Figure~\ref{fig:inv}:
\begin{enumerate}
  \setcounter{enumi}{\value{saveenum}}
\item\label{inv:validity} At any process, if $\phase[m] \not= \Start$, then $m$
  has been previously multicast by a client.

\item\label{inv:ballot-monot} At any process, if $\ballot = b'$ and
  $\aballot = b$, then $b \le b'$ and at any later time at this process 
  $\ballot \ge b'$ and $\aballot \ge b$.


\item\label{inv:ballot-aballot} At any process, we always have
  $\aballot \le \ballot$.

\item\label{inv:phase-monot} At any process, if $\aballot = b$ and
  $\phase[m] = h$, then at any later time at this process, if $\aballot = b$,
  then $\phase \ge h$.

\item\label{inv:clock-monot} At any process, if $\aballot = b$ and
  $\clock = \ts$, then at any later time at this process, if $\aballot = b$,
  then $\clock \ge \ts$.

\item\label{inv:nochange-lts} At any process, if $\aballot = b$,
  $\phase[m] \not= \Start$ and $\localts[m] = \lts$, then at any later time at
  this process, if $\aballot = b$, then $\localts[m] = \lts$.

\item\label{inv:lts-gts} At any process, if $\phase[m] = \Committed$, then
  $\localts[m] \le \globalts[m]$.

\item\label{inv:leader-clock} At any process, if $\phase[m] = \Committed$, then
  $\clock \ge \ttime(\globalts[m])$.

\item\label{inv:prefix} When a process in a group $g_0$ sends
  $\ACCEPTACK(m, g_0, \Bal)$, the $\localts$ at this process is a subset of
  $\localts$ at $\leader(\Bal(g_0))$ when the leader sent the corresponding
  $\ACCEPT$ message.

\item\label{inv:committed} If at a process in $g_0$ we have $\ballot = b'$,
  $\phase[m] = \Committed$, $\localts[m] = \lts$ and $\globalts[m] = \gts$, then
  $\DELIVER(m, b, \lts, \gts)$ has been previously sent to $g_0$ for some
  $b \le b'$.

\item\label{inv:commit2accept} Assume $\DELIVER(m, b, \lts, \gts)$ has been sent
  to a group $g_0$. Then there exist $\Bal(g)$ and $\Lts(g)$ for each
  $g \in \dest(m)$ such that: $\Bal(g_0) \le b$; $\Lts(g_0) = \lts$;
  $\gts = \max\{\Lts(g) \mid g \in \dest(m)\}$; for each $g \in \dest(m)$ a
  quorum of processes in $g$ have received messages
  $\{\ACCEPT(m, g, \Bal(g), \Lts(g)) \mid g \in \dest(m)\}$ and responded to
  them with $\ACCEPTACK(m, g_0, \Bal)$; and the quorum for $g_0$ includes
  $\leader(\Bal(g_0))$.



\item\label{inv:main2} Assume that at some point a set of processes in $g_0$
  have received a message
  $\NEWSTATE(b, \vclock, \vphase, \vlocalts, \vglobalts)$ and responded with
  $\NEWSTATEACK(b)$, and that this set forms a quorum together with
  $\leader(b)$. Assume $\vphase[m] \not= \Start$ and let $\gts = \vglobalts[m]$
  and $\vlocalts_0$ be the projection of $\vlocalts$ to messages $m'$ such that
  $\vphase[m'] \not= \Start \wedge \vlocalts[m'] < \gts$. Whenever at a process
  in $g_0$ we have $\aballot > b$, the condition~(\ref{main-msg}) holds.

\item\label{inv:two-majorities-accept} Assume two quorums of processes in $g_0$
  have respectively received the sets of messages
\begin{gather}
\{\ACCEPT(m, g, \Bal_1(g), \Lts_1(g)) \mid g \in \dest(m)\};\label{quorum-accept1}
\\
\{\ACCEPT(m, g, \Bal_2(g), \Lts_2(g)) \mid g \in \dest(m)\}\label{quorum-accept2}
\end{gather}
and respectively responded to them with
\begin{gather}
\ACCEPTACK(m, g_0, \Bal_1); \label{quorum-acceptack1}
\\
\ACCEPTACK(m, g_0, \Bal_2) \label{quorum-acceptack2}
\end{gather}
Then $\Lts_1(g_0) = \Lts_2(g_0)$.



\item\label{tx-distinct} At any process, the timestamps in the $\localts$ array
  are distinct.

\end{enumerate}

\smallskip
\smallskip

The proofs of Invariants~\ref{inv:unique-lts}
and~\ref{inv:validity}-\ref{inv:prefix} are easy and thus omitted. We now prove
the rest of the invariants.

\paragraph{Proof of Invariant~\ref{inv:committed}.}
We prove the invariant by induction on the length of the execution. Initially no
process has $\phase[m] = \Committed$, so the invariant holds trivially. We next
consider transitions that can affect its validity.

The transition at line~\ref{alg:receive-accept} preserves the invariant due to
Invariant~\ref{inv:nochange-lts}.  The transition at line~\ref{alg:acceptack}
preserves the invariant because of the $\DELIVER$ message it sends. The
transition at line~\ref{alg:receive-deliver} preserves the invariant due to the
check $\ballot = b$ at line~\ref{alg:check-deliver}.

Consider the transition at line~\ref{alg:newleaderack} at a process
$p_i \in g_0$. Assume that after this transition at $p_i$ we have
$\ballot = b'$, $\phase[m] = \Committed$, $\localts[m] = \lts$ and
$\globalts[m] = \gts$.  Then $p_i$ has received a message
$\NEWLEADERACK(b', \_, \_, \vphase, \vlocalts, \vglobalts)$ from some $p_j$,
where $\vphase[m] = \Committed$, $\vlocalts[m] = \lts$ and
$\vglobalts[m] = \gts$. Right before the process $p_j$ sent the $\NEWLEADERACK$
message, it must have had $\ballot < b'$, $\phase[m] = \Committed$,
$\localts[m] = \lts$ and $\globalts[m] = \gts$. Then by the induction hypothesis
$\DELIVER(m, b, \lts, \gts)$ has been sent to $g_0$ for some $b \le b'$, as
required.

It is easy to see that all other transitions trivially preserve (*).\qed

\paragraph{Proof of Invariant~\ref{inv:commit2accept}.} 
We prove the invariant by induction on the length of the execution. Assume the
process $\leader(b)$ sent $\DELIVER(m, b, \lts, \gts)$ to $g_0$. This could have
happened at lines~\ref{alg:send-deliver} or~\ref{alg:send-deliver-recovery}. In
the case of line~\ref{alg:send-deliver} the invariant is preserved by the
precondition of the corresponding transition.  In the case of
line~\ref{alg:send-deliver-recovery}, $\leader(b)$ must have received a message
$\NEWLEADERACK(b, b_0, \_, \vphase, \vlocalts, \vglobalts)$ from a process
$p_j \in g_0$ such that $b_0 < b$, $\vphase[m] = \Committed$,
$\vlocalts[m] = \lts$ and $\vglobalts[m] = \gts$. When $p_j$ sent this message,
it had $\phase[m] = \Committed$, $\localts[m] = \lts$ and $\globalts[m] =
\gts$. Then by Invariant~\ref{inv:committed}, $\DELIVER(m, b_0, \lts, \gts)$ had
been sent to $g_0$. The required then follows from the induction hypothesis.\qed

\paragraph{Proof of Invariants~\ref{inv:main} and~\ref{inv:main-msg}.}
We prove the two invariants together by induction on the value of
$\aballot$. Assume that the invariant holds for all values of $\aballot$ less
than $b'$. We now show it for $\aballot = b'$. Assume that at some point a
quorum $Q$ of processes in $g_0$ have received the set of messages
\begin{equation}\label{main-accept}
\{\ACCEPT(m, g, \Bal(g), \Lts(g)) \mid g \in \dest(m)\}
\end{equation}
and responded to them with 
\begin{equation}\label{main-acceptack}
\ACCEPTACK(m, g_0, \Bal),
\end{equation}
Let $\gts = \max\{\Lts(g) \mid g \in \dest(m)\}$, $\lts = \Lts(g_0)$ and
$b = \Bal(g_0)$. If $\leader(b) \in Q$, let $\vlocalts_0$ be the projection of
$\localts$ when $\leader(\Bal(g_0))$ sent its $\ACCEPTACK$ to messages $m'$ such
that $\phase[m'] \not= \Start \wedge \localts[m'] < \gts$.

We show that, whenever at a process $p_i \in g_0$ we have $\aballot = b' > b$,
the conditions~(\ref{main-phase})-(\ref{main-clock}) hold and
additionally~(\ref{main-msg}) holds if $\leader(b) \in Q$. We prove this
property by induction on the length of the execution. The property holds
trivially at the start of the execution, since at this time
$\aballot = \bot < b$. The only transitions that can affect its validity in a
non-trivial way are those in lines~\ref{alg:receive-accept}
and~\ref{alg:newleaderack}.

Consider the transition by $p_i$ in line~\ref{alg:receive-accept}. This
transition trivially preserves~(\ref{main-phase}) and~(\ref{main-clock}) and
preserves~(\ref{main-lts}) due to Invariant~\ref{inv:nochange-lts}. We now prove
that it preserves~(\ref{main-msg}), provided that $\leader(b) \in Q$. Assume
that after the transition we have $\aballot = b'$ at
$p_i$. Then~(\ref{main-msg}) holds if $p_i$ is a follower, i.e.,
$p_i \not= \leader(b')$, due to Invariant~\ref{inv:prefix} and the fact that by
induction hypothesis~(\ref{main-msg}) holds at $\leader(b')$. Assume now that
$p_i = \leader(b')$. By induction hypothesis, before the transition at this
process we have $\clock \ge \ttime(\gts)$. Then a new local timestamp assigned
by the leader during the transition is higher than $\gts$ and,
thus,~(\ref{main-msg}) is preserved.

Consider now the transition by $p_i$ in line~\ref{alg:newleaderack}. Assume that
after this $\aballot = b'$ at $p_i$. Then $p_i$ must have received messages
$$
\NEWLEADERACK(b', \vaballot(p_j), \vclock(p_j), \vphase(p_j), \vlocalts(p_j),
\vglobalts(p_j)),
$$
from a quorum $Q'$ of processes $p_j$. Let
$b_0 = \max\{\vaballot(p_j) \mid p_j \in Q'\}$ and
$J = \{j \mid \vaballot(p_j) = b_0\}$. Then by the check at
line~\ref{alg:check-newleader} and Invariant~\ref{inv:ballot-aballot}, we have
$b_0 < b'$.

We next establish a couple of auxiliary results. First, we prove that
$J \not= \emptyset$. To this end, note that $Q \cap Q' \not= \emptyset$. Then
some process $p_{j_0} \in Q'$ must have received~(\ref{main-accept}) and
responded with~(\ref{main-acceptack}). The process $p_{j_0}$ must have sent its
$\ACCEPTACK$ message before the $\NEWLEADERACK$ message. By the check in
line~\ref{alg:check-propose}, when $p_{j_0}$ sent the $\ACCEPTACK$ message, it
had $\aballot = b$. Then by Invariant~\ref{inv:ballot-monot}, when $p_{j_0}$
sent its $\NEWLEADERACK$ message, it had $\aballot \ge b$. Hence,
$\vaballot(p_{j_0}) \ge b$ and $b_0 \ge b$, so that $b_0 \not= \bot$ and
$J \not= \emptyset$, as desired.

We next prove the following:
\begin{multline}\label{main-aux4}
  \forall j \in Q'\setminus J.\, \forall m'.\, 
  \vphase[m'] = \Committed \implies
  \\
  \forall j' \in J.\, \vphase(p_{j'})[m'] \ge \Accepted \wedge
  \vlocalts(p_{j'})[m'] = \vlocalts(p_j)[m'].
\end{multline}
To this end, consider an arbitrary $j \in Q' \setminus J$, so that
$\vaballot(p_j) < b_0$.  Further, consider $m'$ such that
$\vphase[m'] = \Committed$ and let $\vlocalts(p_j)[m'] = \lts'$. Right before
sending the $\NEWLEADERACK$ messages, $p_j$ had $\phase[m'] = \Committed$ and
$\localts[m'] = \lts'$. By Invariant~\ref{inv:committed}, a
$\DELIVER(m', b'', \lts', \_)$ message must have been sent for some $b''$ such
that $b'' \le \vaballot(p_j)< b_0 < b'$. Then by
Invariant~\ref{inv:commit2accept} and the induction hypothesis, for any
$j' \in J$ when $p_{j'}$ sent the $\NEWLEADERACK$ message, it must have had
$\phase[m'] \ge \Accepted$ and $\localts[m'] = \lts'$. Then
$\vphase(p_{j'})[m'] \ge \Accepted$ and $\vlocalts(p_{j'})[m'] = \lts'$. We have
thus established~(\ref{main-aux4}).

We now show the desired properties by making a case split on the relationship
between $b_0$ and $b$. Consider first the case when $b_0 = b$. Since
$p_{j_0} \in Q'$ has received and acknowledged~(\ref{main-accept}), by
Invariants~\ref{inv:phase-monot},~\ref{inv:nochange-lts},
and~\ref{inv:clock-monot} we have
\begin{equation}\label{main-aux6}
\vphase(p_{j_0})[m] \ge \Accepted \wedge \vlocalts(p_{j_0})[m]=\lts \wedge
\vclock(p_{j_0}) \ge \ttime(\gts).
\end{equation}
Then right after $p_i$ executes the transition in line~\ref{alg:newleaderack},
properties~(\ref{main-phase}) and~(\ref{main-clock}) hold. Since the leader of a
ballot cannot send messages with different local timestamps for
$m$,~(\ref{main-aux6}) implies
$$
\forall j \in J.\, \vphase(p_j)[m] \ge \Accepted \implies \vlocalts(p_j)[m]=\lts.
$$
Together with~(\ref{main-aux4}) this ensures that condition~(\ref{main-lts})
holds after $p_i$'s transition. 

We now prove~(\ref{main-msg}), provided $\leader(b) \in Q$. Consider arbitrary
$j \in J$ and $m'$ such that $\vphase(p_j)[m'] \ge \Accepted$ and
$\vlocalts(p_{j})[m'] < \gts$. Then $\leader(b)$ sent an $\ACCEPT$ or
$\NEWSTATE$ message carrying the local timestamp $\vlocalts(p_{j})[m']$ for $m'$
to $p_j$. This leader also sent the message $\ACCEPTACK(m, g_0, \Bal)$, after
which the leader has $\clock \ge \ttime(\gts)$. The leader could not issue a
timestamp less than $\gts$ after this and, hence, when the leader sent the
$\ACCEPTACK$ message, it must have already had $\phase[m'] \not= \Start$ and
$\vlocalts_0[m'] = \localts[m'] = \vlocalts(p_{j})[m']$. We have thus
established:
\begin{equation}\label{main-aux1}
\forall j \in J.\, \forall m'.\, \vphase(p_j)[m'] \ge \Accepted \wedge
\vlocalts(p_{j})[m'] < \gts \implies \vlocalts(p_{j})[m'] = \vlocalts_0[m'].
\end{equation}
Together with~(\ref{main-aux4}) this ensures that condition~(\ref{main-msg})
holds after $p_i$'s transition.

Assume now that $b_0 > b$. Consider an arbitrary $j \in J$, so that
$\vaballot(p_{j}) = b_0$. Right before sending the $\NEWLEADERACK$ messages,
$p_j$ had $b < \aballot = b_0 < b'$. Then by the induction hypothesis, $p_j$
also satisfied properties (\ref{main-phase})-(\ref{main-clock}) and, when
$\leader(b) \in Q$, property~(\ref{main-msg}). Hence,
\begin{equation}\label{main-aux2}
\forall j \in J.\,
\vphase(p_{j})[m] \ge \Accepted \wedge \vlocalts(p_{j})[m]=\lts \wedge
\vclock(p_{j}) \ge \ttime(\gts)
\end{equation}
and
\begin{multline}\label{main-aux3}
\leader(b) \in Q \implies \forall j \in J.\, \forall m'.\, 
\vphase(p_j)[m'] \ge \Accepted \wedge
\vlocalts(p_{j})[m'] < \gts \implies
\\
\vlocalts(p_{j})[m'] = \vlocalts_0[m'].
\end{multline}
Then~(\ref{main-aux2}) implies that right after $p_i$ executes the transition in
line~\ref{alg:newleaderack}, properties~(\ref{main-phase})
and~(\ref{main-clock}) hold. We have also
established~(\ref{main-aux4}). Specialising it to $m' = m$, we get
\begin{equation}\label{main-aux5}
  \forall j \in Q'\setminus J.\, \vphase[m] = \Committed \implies
  \\
  \forall j' \in J.\, \vphase(p_{j'})[m] \ge \Accepted \wedge
  \vlocalts(p_{j'})[m] = \lts.
\end{equation}
Together with~(\ref{main-aux2}) this implies that~(\ref{main-lts}) holds.
Finally,~(\ref{main-aux4}) and~(\ref{main-aux3}) imply that~(\ref{main-msg})
holds when $\leader(b) \in Q$.\qed

\paragraph{Proof of Invariant~\ref{inv:main2}.} The proof is analogous to that
of Invariant~\ref{inv:main-msg}.\qed

\paragraph{Proof of Invariant~\ref{inv:two-majorities-accept}.}
We can assume without loss of generality that $\Bal_1(g_0) \le \Bal_2(g_0)$. If
$\Bal_1(g_0) = \Bal_2(g_0)$, then $\leader(\Bal_1(g_0))$ sent
$\ACCEPT(m, g_0, \Bal_1(g_0), \Lts_1(g_0))$ and
$\ACCEPT(m, g_0, \Bal_1(g_0), \Lts_2(g_0))$. Since a leader never assigns
different local timestamps to the same message (line~\ref{alg:lts-exists}), we
must have $\Lts_1(g_0) = \Lts_1(g_0)$. Assume now that
$\Bal_1(g_0) < \Bal_2(g_0)$. Then by Invariant~\ref{inv:main}(a, b) we have
$\phase[m] \ge \Accepted$ and $\localts[m] = \Lts_1(g_0)$ at
$\leader(\Bal_2(g_0))$ when it sends
$\ACCEPT(m, g_0, \Bal_2(g_0), \Lts_2(g_0))$. But then due to the check in
line~\ref{alg:lts-exists} we must have $\Lts_1(g_0) = \Lts_2(g_0)$, as
required. \qed

\paragraph{Proof of Invariant~\ref{inv:consensus}.}  {\bf (a).} Assume that
$\DELIVER(m, \_, \lts_1, \_)$ and $\DELIVER(m, \_, \lts_2, \_)$ have been sent
to the same group $g_0$. By Invariant~\ref{inv:commit2accept}, a quorum of
processes in $g_0$ have received~(\ref{quorum-accept1}) and responded
with~(\ref{quorum-acceptack1}), where $\Lts_1(g_0) = lts_1$. Similarly, a quorum
of processes in $g_0$ have received~(\ref{quorum-accept2}) and responded
with~(\ref{quorum-acceptack2}), where $\Lts_2(g_0) = lts_2$. Then by
Invariant~\ref{inv:two-majorities-accept} we have $\lts_1 = \lts_2$.

{\bf (b).} Assume that $\DELIVER(m, \_, \_, \gts_1)$ and
$\DELIVER(m, \_, \_, \gts_2)$ have been sent and $\gts_1 \not= \gts_2$. Then by
Invariant~\ref{inv:commit2accept} for some group $g_0$,
messages~(\ref{quorum-accept1}) and~(\ref{quorum-accept2}) have been sent to
$g_0$ and acknowledged by quorums, and furthermore,
$\Lts_1(g_0) \not= \Lts_2(g_0)$. But this contradicts
Invariant~\ref{inv:two-majorities-accept}. \qed

\paragraph{Proof of Invariant~\ref{tx-distinct}.} We prove the invariant by
induction on the length of the execution. The updates of $\localts$ at a process
with $\status = \LEADER$ preserve the invariant because the leader always issues
fresh local timestamps to new messages (the handler at
line~\ref{alg:leader-origin}). Assume that $\localts$ is updated at a process
with $\status = \FOLLOWER$ because the process receives a message from the
leader of its ballot. By Invariant~\ref{inv:prefix}, the $\localts$ at the
follower is a subset of the $\localts$ at the leader when it sent the
message. By the induction hypothesis, all entries in $\localts$ at the leader
are distinct when the leader sends the message and, hence, they are also
distinct at the follower after it receives the message. We next prove that
updates to $\localts$ preserve the invariant when the process has
$\status = \RECOVERING$. The transition at line~\ref{alg:trans-newleadersync}
trivially preserves the invariant by the induction hypothesis. It remains to
prove that the invariant is preserved by the transition at
line~\ref{alg:newleaderack}.

Consider such a transition at a process $p_i$ that receives $\NEWLEADERACK$
messages from a quorum $Q'$. Let $b_0 = \max\{\vaballot(p_j) \mid p_j \in Q'\}$,
so that $J = \{j \mid \vaballot(p_j) = b_0\}$. Consider two messages $m_1$ and
$m_2$ such that $m_1 \not= m_2$ and assume $j_1, j_2 \in J$ are such that
$\vphase(p_{j_1})[m_1] \ge \Accepted$ and $\vphase(p_{j_2})[m_2] \ge
\Accepted$. Let $\vlocalts(p_{j_1})[m_1] = \lts_1$ and
$\vlocalts(p_{j_2})[m_2] = \lts_2$. Then when the processes $p_{j_1}$ and
$p_{j_2}$ sent the $\NEWLEADERACK$ messages, they respectively had
$$
\phase[m_1] \ge \Accepted,\ 
\localts[m_1] = \lts_1,\ 
\aballot = b_0 \  \ \mbox{at $p_{j_1}$};
$$
$$
\phase[m_2] \ge \Accepted,\ 
\localts[m_2] = \lts_2,\ 
\aballot = b_0 \ \ \mbox{at $p_{j_2}$}.
$$
These processes must have received the timestamps $\lts_1$ and $\lts_2$ in
appropriate messages sent by $\leader(b_0)$. When the leader sent the last of
these messages, it had $\localts[m_1] = \lts_1$ and $\localts[m_2] = \lts_2$. By
induction hypothesis, $\lts_1 \not= \lts_2$. We have thus proved
\begin{multline*}
  \forall m_1, m_2.\, \forall j_1, j_2 \in J.\, m_1 \not= m_2 \wedge
  \vphase(p_{j_1})[m_1] \ge \Accepted \wedge \vphase(p_{j_2})[m_2] \ge \Accepted
  \implies
  \\
  \vlocalts(p_{j_1})[m_1] \not= \vlocalts(p_{j_2})[m_2].
\end{multline*}
Like in the proof of Invariant~\ref{inv:main}, we can also
establish~(\ref{main-aux4}). Together with the above, this implies that at the
end of the transition at line~\ref{alg:newleaderack} all entries in $\localts$
are distinct.\qed

\paragraph{Proof of Invariant~\ref{inv:unique-gts}.} 
We prove the required by contradiction. Assume $\DELIVER(m_1, \_, \_, \gts)$ and
$\DELIVER(m_2, \_, \_, \gts)$ are sent and $m_1 \not= m_2$. Let
$\gts = (t, g_0)$. Then $g_0 \in \dest(m_1) \cap \dest(m_2)$. By
Invariant~\ref{inv:commit2accept}, a quorum of processes in $g_0$ have received
$$
\{\ACCEPT(m_1, g, \Bal_1(g), \Lts_1(g)) \mid g \in \dest(m_1)\}
$$
and responded with
$$
\ACCEPTACK(m_1, g_0, \Bal_1),
$$
and $\gts = \Lts_1(g_0)$. Analogously, a quorum of
processes in $g_0$ have received
$$
\{\ACCEPT(m_2, g, \Bal_2(g), \Lts_2(g)) \mid g \in \dest(m_2)\}
$$
and responded with
$$
\ACCEPTACK(m_1, g_0, \Bal_2),
$$
and $\gts = \Lts_2(g_0)$. Hence, $\Lts_1(g_0) = \gts = \Lts_2(g_0)$. Without
loss of generality, we assume $\Bal_1(g_0) \le \Bal_2(g_0)$. If
$\Bal_1(g_0) = \Bal_2(g_0) = b$, then $\leader(b)$ sent the messages
$\ACCEPT(m_1, g_0, b, \gts)$ and $\ACCEPT(m_2, g_0, b, \gts)$. But this is
impossible due to Invariant~\ref{tx-distinct} and the fact that a leader
advances $\clock$ when assigning a new local timestamp
(line~\ref{alg:clock-inc}). Hence, $\Bal_1(g_0) < \Bal_2(g_0)$. Then by
Invariant~\ref{inv:main}(b), when $\leader(\Bal_2(g_0))$ sends
$\ACCEPT(m_2, g_0, \Bal_2(g_0), \gts)$, it has $\localts[m_1] = \gts$. But this
contradicts Invariant~\ref{tx-distinct}.  \qed

\paragraph{Proof of Theorem~\ref{thm:correctness} (Validity, Integrity, Ordering).}
Validity follows from Invariant~\ref{inv:validity}. Integrity follows from
the update in line~\ref{alg:bump-last-gts}, the check in
line~\ref{alg:check-deliver} and Invariant~\ref{inv:consensus}.

It remains to prove Ordering. Given a run of the protocol, we define a
relation $\sqsubset$ on messages as follows: $m \sqsubset m'$ if there exists a
group $g \in \dest(m) \cap \dest(m')$ and a process $p \in g$ such that $p$
delivers $m$ before delivering $m'$. We next prove that $\sqsubset$ is
acyclic. Then the relation $\prec$ required by (Ordering) can be constructed as
any total order containing $\sqsubset$.

For a client message $m$ to be delivered, a message
$\DELIVER(m, \_, \_, \gts_m)$ must have been sent for some timestamp
$\gts_m$. By Invariants~\ref{inv:consensus} and~\ref{inv:unique-gts}, such a
$\gts_m$ is well-defined and unique to a given $m$. If a client message $m$ has
not been delivered at any process, we let $\gts_m = \top$, where $\top$ is
higher than all other timestamps. We now prove that
$m_1 \sqsubset m_2 \implies \gts_{m_1} < \gts_{m_2}$, which implies the acyclicity
of $\sqsubset$. To this end, we prove the contrapositive:
$\gts_{m_2} < \gts_{m_1} \implies \neg(m_1 \sqsubset m_2)$. Let
$\gts_{m_1} = \gts_1$ and $\gts_{m_2} = \gts_2$ and assume $\gts_2 <
\gts_1$. Assume that a process $p_i$ in $g_0$ delivers $m_1$ and that
$g_0 \in \dest(m_2)$. We need to prove that $p_i$ has already delivered $m_2$.

The process $p_i$ must deliver $m_1$ due to receiving a message
$\DELIVER(m_1, b_1, \_, \gts_1)$ for some $b_1$. Since $\gts_2 < \gts_1$, we
have $\gts_2 < \top$, and hence a message $\DELIVER(m_2, \_, \_, \gts_2)$ has
been sent in some group. Then by Invariant~\ref{inv:commit2accept} a quorum in
group $g_0$, including $\leader(\Bal_2(g_0))$, have received
$$
\{\ACCEPT(m_2, g, \Bal_2(g), \Lts_2(g)) \mid g \in \dest(m_2)\}
$$
and responded with
$$
\ACCEPTACK(m_2, g_0, \Bal_2),
$$
and $\gts_2 = \max\{\Lts_2(g) \mid g \in \dest(m_2)\}$. Since $\gts_2 < \gts_1$,
we have $\Lts_2(g_0) < \gts_1$. We now make a case split on the relationship
between $b_1$ and $\Bal_2(g_0)$.
\begin{itemize}
\item $b_1 = \Bal_2(g_0)$. Then $\leader(b_1)$ sends
  $\DELIVER(m_1, b_1, \_, \gts_1)$ and $\ACCEPT(m_2, g_0, b_1, \Lts_2(g_0))$. By
  Invariant~\ref{inv:leader-clock}, when $\leader(b_1)$ sends the $\DELIVER$
  message, it has $\clock \ge \ttime(\gts_1)$. If the $\ACCEPT$ message was sent
  after the $\DELIVER$ message, it could not result from the leader receiving
  $m_2$ for the first time: in this case we would have $\Lts_2(g_0) >
  \gts_1$. Hence, when the leader sent the $\DELIVER$ message, it had
  $\phase[m_2] \not= \Start$ and $\localts[m_2] = \Lts_2(g_0)$. Since
  $\Lts_2(g_0) < \gts_1$, due to the checks in lines~\ref{alg:deliv-check1}
  and~\ref{alg:deliv-check2}, at this moment the leader had
  $\phase[m_2] = \Committed$. By Invariants~\ref{inv:committed}
  and~\ref{inv:consensus} the leader also had $\globalts[m_2] = \gts_2$.  Then
  the leader must have sent $\DELIVER(m_2, b_1, \_, \gts_2)$ before
  $\DELIVER(m_1, b_1, \_, \gts_1)$, so that $m_2$ had to be delivered before
  $m_1$.
\item $\Bal_2(g_0) < b_1$. By Invariant~\ref{inv:main}(a, b), when $\leader(b_1)$
  sends the message $\DELIVER(m_1, b_1, \_, \gts_1)$, it has
  $\phase[m_2] \ge \Accepted$ and $\localts[m_2] = \Lts_2(g_0)$. The proof is
  completed as in the previous case.
\item $b_1 < \Bal_2(g_0)$ and the $\DELIVER(m_1, b_1, \_, \gts_1)$ message is
  sent at line~\ref{alg:send-deliver}. Then a quorum in group $g_0$, including
  $\leader(b_1)$, have received
$$
\{\ACCEPT(m_1, g, \Bal_1(g), \Lts_1(g)) \mid g \in \dest(m_2)\}
$$
and responded with
$$
\ACCEPTACK(m_1, g_0, \Bal_1);
$$
furthermore, $\gts_1 = \max\{\Lts_1(g) \mid g \in \dest(m_1)\}$ and
$\Bal_1(g_0) = b_1$. When $\leader(\Bal_2(g_0))$ sent
$\ACCEPT(m_2, g_0, \Bal_2(g_0), \Lts_2(g_0))$, it had $\phase[m_2] \not= \Start$
and $\localts[m_2] = \Lts_2(g_0) < \gts_1$. Since $b_1 < \Bal_2(g_0)$, by
Invariant~\ref{inv:main-msg}, when $\leader(b_1)$ sent its $\ACCEPTACK$ message,
it also had $\phase[m_2] \not= \Start$ and $\localts[m_2] = \Lts_2(g_0)$. Since
$\Lts_2(g_0) < \gts_1$, this must have also been true when $\leader(b_1)$ sent
the $\DELIVER(m_1, b_1, \_, \gts_1)$, since after this the leader has
$\clock \ge \ttime(\gts_1)$ and thus cannot issue local timestamps lower than
$\gts_1$. At this moment the leader had $\phase[m_1] = \Committed$ and
$\globalts[m_1] = \gts_1$. Then due to the check in line~\ref{alg:deliv-check1},
the leader should also have $\phase[m_2] = \Committed$ and, by
Invariants~\ref{inv:committed} and~\ref{inv:consensus},
$\globalts[m_2] = \gts_2$. But then the leader would have to deliver $m_2$
before $m_1$, as required.

\item $b_1 < \Bal_2(g_0)$ and the $\DELIVER(m_1, b_1, \_, \gts_1)$ message is
  sent at line~\ref{alg:send-deliver-recovery}. This case is handled similarly
  to the previous one, but using Invariant~\ref{inv:main2} instead of
  Invariant~\ref{inv:main-msg}.
\end{itemize}\qed


\bigskip

We next prove Termination.
For an application message $m$, let  $\Q_m$ denote the set of all
correct quorums of every group in $\dest(m)$, and $\Leads_m$ denote the set of the
leaders of the quorums in $\Q_m$ as stipulated by
Invariant~\ref{inv:main-live} in Figure~\ref{fig:inv}.
We first prove the following auxiliary lemmas.
\smallskip
\smallskip
\begin{lemma}
Let $m$ be an application message and $t$ be the time
stipulated by Invariant~\ref{inv:main-live} in Figure~\ref{fig:inv}. 
Suppose that all processes in $\Leads_m$ receive $\MULTICAST(m)$ after $t$. Then 
eventually $m$ is committed at all processes in $\Leads_m$.
\label{lem:mcast-commit}
\end{lemma}

\paragraph{Proof.}
Since each process $p_j\in \Leads_m$ receives $\MULTICAST(m)$ at time $t^1_j > t$, 
Invariant~\ref{inv:main-live} in Figure~\ref{fig:inv} implies 
that $p_j$ considers itself a leader at $t^1_j$, validating the guard
in line~\ref{alg:status-leader} of Figure~\ref{fig:protocol}.
This causes $p_j$ to execute the code in line~\ref{alg:send-accept} 
of Figure~\ref{fig:protocol} 
causing it to send $\ACCEPT(m, \_, \_, \_)$ messages 
to the members of all groups in $\dest(m)$. Since the channels are reliable,
all correct processes in all groups in $\dest(m)$ will eventually
receive $\ACCEPT(m, \_, \_, \_)$ messages from all processes
$p_j \in \Leads_m$. Since every group in $\dest(m)$ includes a quorum
of correct processes that follows its leader, the guard in 
line~\ref{alg:check-propose} of Figure~\ref{fig:protocol} will 
hold at all processes in all quorums in $\Q_m$ causing them to 
eventually respond with $\ACCEPTACK(m, \_, \_)$
to $p_j$. 
Since the channels are reliable, there exists a time $t^2_j \ge t^1_j$
at which $p_j$ will have received  $\ACCEPTACK(m, \_, \_)$
messages from all quorums in $\Q_m$, and 
$\ACCEPT(m, \_, \_, \_)$ from all leaders in $\Leads_m$.
Furthermore, since $t^2_j > t$, $p_j$ considers itself a leader
and its $\aballot$ has the same value as it had when it sent 
$\ACCEPT(m, \_, \_, \_)$. This implies that the guard
in line~\ref{alg:pre-acceptack} of Figure~\ref{fig:protocol}
is true at $t^2_j$, enabling $p_j$ to reach line~\ref{alg:deliv-check1}
in which $m$ is committed.
\qed
\smallskip
\smallskip
\begin{lemma}
Let $m$ be an application message and $t$ be the time
stipulated by Invariant~\ref{inv:main-live} in Figure~\ref{fig:inv}.
If a process in $\Leads_m$ invokes $\retry(m)$ after $t$, then
eventually all processes in $\Leads_m$ receive $\MULTICAST(m)$.
\label{lem:retry}
\end{lemma}

\paragraph{Proof.}
Suppose that a process $p_i\in \Leads_m$ invokes $\retry(m)$
at time $t_r > t$. Then
$p_i$ will send $\MULTICAST(m)$ messages to all processes in the set 
$\{\currleaders[g'] \mid g'\in \dest(m)\}$. 
Since $t_r > t$,  by Invariant~\ref{inv:main-live}
in Figure~\ref{fig:inv} this set coincides with $\Leads_m$. Given that 
the channels are reliable, this implies that all processes in $\Leads_m$
will eventually receive $\MULTICAST(m)$ after $t_r > t$.
\qed
\smallskip
\smallskip
\begin{lemma}
Let $m$ be an application message and $t$ be the time
stipulated by Invariant~\ref{inv:main-live} in Figure~\ref{fig:inv}.
If a process in $\Leads_m$ has $\phase[m]\in \{\Proposed, \Accepted\}$
at time $t' \ge t$, then eventually all processes
in $\Leads_m$ receive $\MULTICAST(m)$.
\label{lem:prop-accept}
\end{lemma}

\paragraph{Proof.}
Consider $p_i\in \Leads_m$, and suppose that 
$\phase[m] \in \{\Proposed, \Accepted\}$  at $p_i$ at
time $t' \ge t$. 
By message recovery  mechanism (\S\ref{sec:optimised}),  there exists
a time $t_r > t$ such that  either $m$ is committed  at $p_i$ at $t_r$, or 
$p_i$ invokes $\retry(m)$ at $t_r$.  If the latter occurs, then by
Lemma~\ref{lem:retry}, all processes in $\Leads_m$ eventually receive
$\MULTICAST(m)$.
\qed

\paragraph{Proof of Lemma~\ref{lem:all-commit}.}
Let $p_i\in \Leads_m$, and suppose that $m$ is known to $p_i$, but not yet
committed at $t$. Then $\phase[m] \in \{\Proposed, \Accepted\}$ at $p_i$ at
$t$. By Lemma~\ref{lem:prop-accept}, all processes in $\Leads_m$ eventually
receive $\MULTICAST(m)$. Thus, by Lemma~\ref{lem:mcast-commit}, $m$ is
eventually committed at $p_i$. Next, suppose that $m$ is received by $p_i$ via
$\MULTICAST(m)$ at $t' > t$. Since $t'>t$, by Invariant~\ref{inv:main-live} in
Figure~\ref{fig:inv}, $p_i$ considers itself a leader at $t'$. Thus, the guard
in line~\ref{alg:status-leader} of Figure~\ref{fig:protocol} holds, enabling
$p_i$ to execute the code in
lines~\ref{alg:lts-exists}-\ref{alg:set-phase-proposed} of
Figure~\ref{fig:protocol}. This implies that upon reaching
line~\ref{alg:send-accept}, $m$ is either committed at $p_i$ or
$\phase[m]\in \{\Proposed, \Accepted\}$. If the latter holds, then by
Lemma~\ref{lem:prop-accept}, all processes in $\Leads_m$ eventually receive
$\MULTICAST(m)$. Thus, by Lemma~\ref{lem:mcast-commit}, $m$ is eventually
committed at $p_i$.  \qed


\smallskip
\smallskip
\begin{lemma}
Let $m$ be an application message, and suppose that
a correct process $p_i$ invokes $\multicast(m)$. Then $m$ is eventually
committed at all processes in $\Leads_m$.
\label{lem:correct-mult}
\end{lemma}

\paragraph{Proof.}
By the message recovery mechanism (\S\ref{sec:optimised}), $p_i$
will continue retransmitting $\MULTICAST(m)$ to the processes
in $\{\currleaders[g] \mid g\in \dest(m)\}$ until  $m$ is committed
at the leader of some group $g\in \dest(m)$. 
Since by Invariant~\ref{inv:main-live}
in Figure~\ref{fig:inv}, $\Leads_m = \{\currleaders[g] \mid g\in \dest(m)\}$
after $t$, there exists time $t'$ such that
either \emph{(i)} $\MULTICAST(m)$ issued by one of the retransmission attempts 
reaches all processes in $\Leads_m$ after $t$, or 
\emph{(ii)} $m$ is committed by the leader of some 
group $g\in \dest(m)$. 
If \emph{(i)} holds, then by Lemma~\ref{lem:mcast-commit}, all processes
in $\Leads_m$ eventually commit $m$. If \emph{(ii)} holds, then there exists
a quorum $Q$ of processes in every destination group $g'$ of $m$ that
received $\ACCEPT(m, \_, \_, \_)$ from the leader of $g'$, and responded
to it with $\ACCEPTACK(m, \_, \_, \_)$. 
By Invariant~\ref{main-phase} in Figure~\ref{fig:inv},
this implies that $\phase[m] \in \{\Accepted,\Committed\}$
at all processes in $\Leads_m$ starting from some time $t'' \ge t$. 
Suppose that there exists a process $p_j\in \Leads_m$ 
such that $\phase[m]=\Accepted$ at $p_j$ at $t''$. 
By Lemma~\ref{lem:prop-accept}, this implies that 
all processes in $\Leads_m$ eventually receive
$\MULTICAST(m)$. Thus,  by Lemma~\ref{lem:mcast-commit}, $m$ is eventually
committed at all processes in $\Leads_m$.
\qed
\smallskip
\smallskip
\begin{lemma}
Let $m$ be an application message, and suppose that
some process delivers $m$. Then $m$ is eventually
committed at all processes in $\Leads_m$.
\label{lem:correct-delivery}
\end{lemma}

\paragraph{Proof.}
Since some process delivers $m$, $m$ must have been committed at some leader
$p_i$ prior to $m$ being delivered. This implies that
there exists
a quorum $Q$ of processes in every destination group $g'$ of $m$ that
received $\ACCEPT(m, \_, \_, \_)$ from the leader of $g'$, and responded
to it with $\ACCEPTACK(m, \_, \_, \_)$. 
By Invariant~\ref{main-phase} in Figure~\ref{fig:inv},
this implies that $\phase[m] \in \{\Accepted,\Committed\}$
at all processes in $\Leads_m$ starting from some time $t'' \ge t$. 
Suppose that there exists a process $p_j\in \Leads_m$ 
such that $\phase[m]=\Accepted$ at $p_j$ at $t''$. 
By Lemma~\ref{lem:prop-accept}, this implies that 
all processes in $\Leads_m$ eventually receive
$\MULTICAST(m)$. Thus,  by Lemma~\ref{lem:mcast-commit}, $m$ is eventually
committed at all processes in $\Leads_m$.
\qed

\paragraph{Proof of Theorem~\ref{thm:correctness} (Termination).}
Let $m$ be an application message and suppose that $m$ is either
multicast by a correct process, or delivered by any process.
Then by Lemmas~\ref{lem:correct-mult} and \ref{lem:correct-delivery},
$m$ is eventually committed at all processes in $\Leads_m$. 
Consider a group $g\in \dest(m)$, and let $Q$ be the corresponding
quorum of $g$ in $\Q_m$ and $p_i$ be the leader of $Q$ in 
$\Leads_m$. Let $t_i$ be the time at which $p_i$ receives
$\NEWSTATEACK$ messages from all processes in $Q$ and becomes
the permanent leader of $Q$, and $t_c \ge t_i$ be the earliest
time at which $\phase[m]=\Committed$ at $p_i$ at or after $t_i$.
By the code in lines~\ref{alg:deliv-check2}-\ref{alg:send-deliver-recovery}, and
lines~\ref{alg:deliv-check1}-\ref{alg:send-deliver} of Figure~\ref{fig:protocol},
$p_i$ will deliver $m$ and send $\DELIVER(m)$ to all members of 
$Q$ at $t_c$ unless it has some uncommitted messages 
whose local timestamps are lower than the one stored 
$\globalts[m]$ at $t_c$. Since 
the $p_i$'s $\clock$ is set above $\globalts[m]$ after $m$ is committed,
all future messages $m'$ received by $p_i$ via $\MULTICAST(m')$ will have 
$\localts[m'] > \globalts[m]$. Thus, there can ever be only finitely many
uncommitted messages $m''$ at $p_i$ such that 
$\localts[m''] < \globalts[m]$. Since $t \ge t_i$, all these
messages are either known to $p_i$ at $t$, or will be received 
via $\MULTICAST$ after $t$, and therefore, by Lemma~\ref{lem:all-commit},
will eventually commit at $p_i$. Once this happens, $p_i$ will deliver
$m$ and send $\DELIVER(m)$ to all members of $Q$. Thus, we conclude 
that $p_i$ delivers $m$ and sends $\DELIVER(m)$ to all members of $Q$
at some point after $t_c$. Since the channels are reliable and 
all members of $Q$ (including $p_i$) are correct, all processes
in $Q$ will eventually deliver $m$, validating the claim.
\qed

\clearpage

\section{Additional Details about Experiments}
\label{app:exp}

\begin{figure*}[h]
\includegraphics[width=\textwidth]{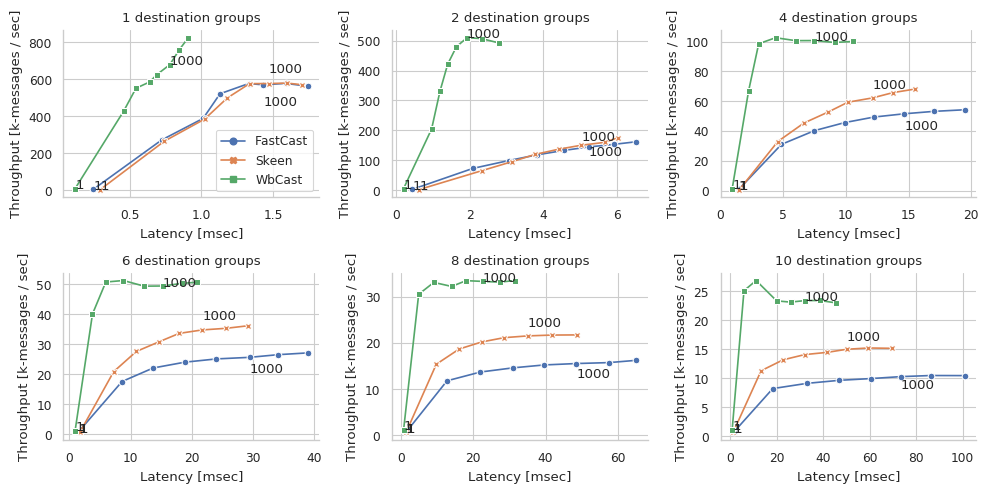}
\caption{Extended version of Figure~\ref{fig:graph-lan}.}
\label{fig:graph-lan-big}
\end{figure*}

\begin{figure*}[h]
\includegraphics[width=\textwidth]{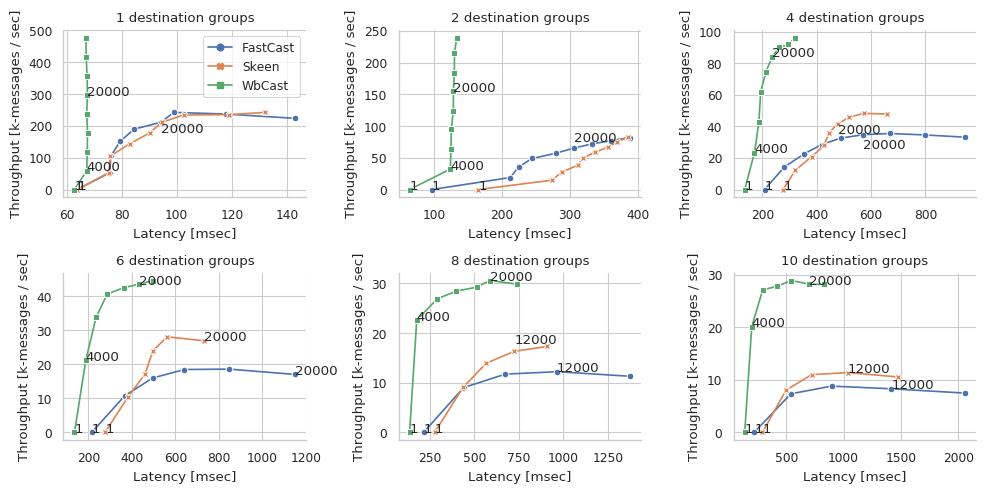}
\caption{Extended version of Figure~\ref{fig:graph-wan}.}
\label{fig:graph-wan-big}
\end{figure*}

\begin{figure*}[t]
\includegraphics[width=\textwidth]{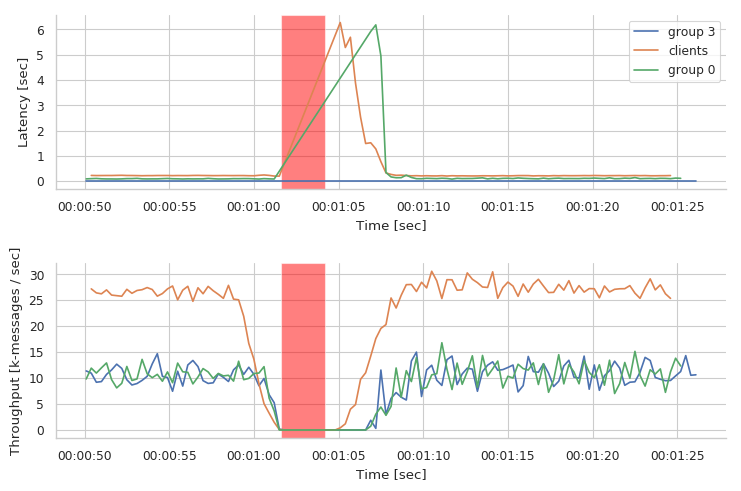}
\caption{Performance of white-box multicast in WAN when the leader of group 3
  crashes. In this experiment, 6000 client threads hosted on 9 machines
  scattered across 3 data centres multicast messages to subsets of 4 out of 10
  groups. We consider the latency and throughput at selected group leaders and
  the aggregate of all client threads, computed in 0.3 sec time bins. We mark in
  red the time frame during which the new leader is being established.}
\label{fig:graph-recovery}
\end{figure*}


\fi

\end{document}